\documentclass[review,5p,times,twocolumn]{elsarticle}
\usepackage[utf8]{inputenc}
\usepackage{graphicx}
\usepackage{algorithm ,algorithmic}
\usepackage{varwidth}
\usepackage[english]{babel}
\usepackage{amssymb, amsthm, amsmath}
\usepackage{mathtools}
\usepackage{array}
\usepackage{rotating}
\usepackage{float}
\usepackage{multirow}
\usepackage{stmaryrd}
\usepackage{booktabs}
\usepackage[font=small,labelfont=bf]{caption}
\usepackage{subcaption}
\usepackage{color}
\usepackage{epsfig}
\usepackage{colortbl}
\usepackage[table]{xcolor}
\usepackage{hhline}
\usepackage[autolanguage]{numprint}
\usepackage{todonotes}

\theoremstyle{definition}
\newtheorem{definition}{Definition}[section]

\definecolor{lightgray}{gray}{0.9}

\begin{document}

\begin{frontmatter}

\title{Efficient anytime algorithms to solve the bi-objective Next Release Problem}

\author[1]{Miguel Ángel Domínguez-Ríos\corref{cor1}}
\ead{miguel.angel.dominguez.rios@uma.es}

\author[1]{Francisco Chicano}
\ead{chicano@lcc.uma.es}

\author[1]{Enrique Alba}
\ead{eat@lcc.uma.es}

\author[2]{Isabel del Águila}
\ead{imaguila@ual.es}

\author[2]{José del Sagrado}
\ead{jsagrado@ual.es}

\address[1]{Dept. Lenguajes y Ciencias de la Computación. Universidad de Málaga (Spain)}
\address[2] {Dept. Informática. Universidad de Almería (Spain)}

\cortext[cor1]{Corresponding author}

%\maketitle

\begin{abstract}
The Next Release Problem consists in selecting a subset of  requirements to develop in the next release of a software product. The selection should be done in a way that maximizes the satisfaction of the stakeholders while the development cost is minimized and the constraints of the requirements are fulfilled. Recent works have solved the problem using exact methods based on Integer Linear Programming. In practice, there is no need to compute all the efficient solutions of the problem; a well-spread set in the objective space is more convenient for the decision maker. The exact methods used in the past to find the complete Pareto front explore the objective space in a lexicographic order or use a weighted sum of the objectives to solve a single-objective problem, finding only supported solutions. In this work, we propose five new methods that maintain a well-spread set of solutions at any time during the search, so that the decision maker can stop the algorithm when a large enough set of solutions is found. The methods are called anytime due to this feature. They find both supported and non-supported solutions, and can complete the whole Pareto front if the time provided is long enough. 

\end{abstract}

\begin{keyword}
Next Release Problem \sep
Multi-objective optimization \sep
Search-based software engineering \sep
Anytime algorithm \sep
Pareto front
\end{keyword}

\end{frontmatter}

\section{Introduction}

Developing software systems that meet stakeholders’ needs and expectations is the ultimate goal of any software provider seeking a competitive edge. Software development processes, specially if agile methodologies are applied, carry requirements prioritization, not only as a way to identify and filter the important requirements, but also to solve conflicts and plan the different releases or deliveries of the software product~\cite{Karlsson1997RE}. These complex decisions require a detailed knowledge of the domain and good quantification and estimation techniques of the requirements' properties that usually involve contradictory criteria. These criteria are defined either by the customers or product owners (e.g. requirement value, deadline), or by the developers (e.g. available effort, team size), or maybe by both of them (e.g. risk, volatility), or even from external factors such as market opportunity.
Release related decisions pose a challenging problem because of their complexity and
dependencies, the number of stakeholders involved in them, the variety of variables that need to be reviewed, and the uncertainty of the information it is relying upon~\cite{Ruhe2010}.

Some authors differentiate  between several kinds of release decisions~\cite{Li2007,Li2010,Ruhe2003}, those related to find the best combination of features to implement in a sequence of releases (i.e. releases schedule) and others that focus on finding the optimal combination of requirements only for the next release, as we do in this paper. Nevertheless, requirements change drastically and frequently. Thus, the company needs a streamlined, flexible approach to release planning. Priorities may shift as the context evolves and as more information becomes available.  As the requirements are further refined, requirements selection is done at a more granular level and will incorporate additional bases when they become appropriate.

Due to the computational complexity of the problem of choosing a set of requirements that will add the most to a software product, it has been also formulated as an optimization problem. Search methods emerged as an alternative strategy to solve it within the framework defined by Search Based Software Engineering (SBSE) discipline~\cite{Harman2001, Harman2014}. This discipline has been successfully and prolifically applied to different problems in requirements engineering (e.g. requirements prioritization, requirements selection, release planning, next release problem, requirement triage)~\cite{pitangueira2015}. In most of the literature, Metaheuristic  algorithms have been applied to solve these problems. 

In a recent work, Veerapen et al.~\cite{veerapen2015integer} showed that Integer Linear Programming solvers can, nowadays, solve the bi-objective version of the Next Release Problem in a few hours for reasonable sizes of the instances. They used the $\varepsilon$-\emph{constraint} method to find the complete Pareto front of the problem. 

Finding the whole Pareto front might require too much computational time in practice, even hours or days. For example, this happens in instances with many non-dominated points. The drawback of $\varepsilon$-constraint is that if the algorithm is stopped before it finishes, the partial Pareto front could lie in a specific region because it finds the solutions in lexicographic order according to some objective, and, therefore, it could be useless to the decision maker. From a practical point of view, the decision maker should be interested in a set of solutions as well spread as possible in the objective space. This is achieved by designing algorithms which \emph{jump} in the objective space, finding scattered solutions and getting the whole Pareto front if there is enough available time. These algorithms are known as \emph{anytime}, because the decision maker can interrupt the execution whenever s/he wishes, and take the partial Pareto front provided by the algorithm. Veerapen et al. used the dichotomic search~\cite{UlunguTeghem1995} to solve the bi-objective NRP. The dichotomic search can only find supported solutions, missing the non-supported ones. 
In all of the instances used in the work of Veerapen et al.~\cite{veerapen2015integer}, the number of supported efficient solutions is below 6\% of the total number of efficient solutions.

In this work, we improve the state-of-the-art methods for solving the bi-objective Next Release Problem by defining five new anytime algorithms for solving bi-objective optimization problems and we apply them to the problem. Four of the designed algorithms are able to find all the efficient solutions, not only the supported ones, and can provide a well-spread set of efficient solutions in a few seconds. Thus, they are more appropriate than the previous state-of-the-art techniques when a set of non-dominated solutions is required in a short time. This work answers the following two main Research Questions:
% Research questions
\begin{itemize}
\item[RQ1] Which of the proposed anytime algorithms is the best one applied to the bi-objective Next Release Problem?
\item[RQ2] Do anytime algorithms find a better-spread set of solutions in the objective space than the classical algorithms when there is a time limit?
\end{itemize}

The paper is organized as follows. The formulation of the bi-objective Next Release Problem is presented in   Section~\ref{sec:problem_definition}. In Section~\ref{sec:Background}, we define concepts related to multi-objective integer linear problems. In Section~\ref{sec:anytime_methods}, five different anytime algorithms are described to solve the bi-objective Next Release Problem. Section~\ref{sec:computational_results} presents the analysis of the algorithms and the computational results. Section~\ref{sec:discussion} presents a discussion on the utility of the new anytime algorithms from the point of view of requirements engineering, while Section~\ref{sec:threats} describes the identified threats to validity. Section~\ref{sec:review} analyzes the relevant previous work on Next Release Problem. The last section, presents our final conclusions and future work.

\section{Next Release Problem formulation}
\label{sec:problem_definition}

The Next Release Problem (NRP) was originally proposed by Bagnall et al.~\cite{bagnall2001next}. It consists in finding a subset of requirements or a subset of stakeholders that maximizes a desirable property, such as revenue, while being constrained by an upper bound on the cost. The bi-objective NRP was formulated by Zhang et al.~\cite{zhang2007multi}. In this case, the upper bound of the cost is lifted and that constraint is transformed into a second objective. Then, the decision-maker is presented with a set of solutions which are all efficient in the Pareto sense.

Let $R$ be the set of $n$ requirements which are not developed yet and $r=(r_{1},...,r_{n}) \in \{0,1\}^{n}$ the binary vector of requirements, where the component $r_i$ takes the value 1 if and only if the $i$-th requirement will be selected for the next release. Let $S$ be the set of $m$ stakeholders and  $s=(s_{1},...,s_{m}) \in \{0,1\}^{m}$ the binary stakeholders vector, where the $k$-th component is set to 1 if and only if the requirements of stakeholder $k$ are included in the next release. Let $c=(c_{1},...,c_{n})$ be the cost vector associated to the requirements and $w=(w_{1},...,w_{m})$ the weight vector associated to the stakeholders, which represents the importance of each stakeholder. To define the constraints, let $P$ be the set of pairs $(i,j)$ where requirement $i$ is a prerequisite for requirement $j$ and let $Q$ be the set of pairs $(i,k)$ where requirement $i$ is requested by stakeholder $k$.

The bi-objective NRP is formulated as:
\begin{align}
\label{eqn:nrp-objs} & \min f(x) = \left( f_{1}(s) = - \sum_{k1=1}^{m} w_{k}s_{k} \quad , \quad  f_{2}(r) = \sum_{i=1}^{n} c_{i}r_{i} \right)\\
& \textit{subject to} \notag \\
\label{eqn:precedence-constraint} & \qquad r_{i} \geq r_{j} \quad \forall (i,j) \in P \\
\label{eqn:stakeholders-constraint} & \qquad r_{i} \geq s_{k} \quad \forall (i,k) \in Q \\
\label{eqn:req-domain} & \qquad r_i \in \{0,1\} \quad \forall i \in 1, \ldots, n\\
\label{eqn:stakeholders-domain} & \qquad s_k \in \{0,1\} \quad \forall k \in 1, \ldots, m
\end{align}
where Eq.~\eqref{eqn:nrp-objs} are the two objective functions to be minimized (satisfaction is to be maximized and this is why it is preceded by a minus sign), Eq.~\eqref{eqn:precedence-constraint} are the precedence constraints among the requirements, Eq.~\eqref{eqn:stakeholders-constraint} forces all requirements of a stakeholder to be implemented in order to satisfy him/her, and Eqs.~\eqref{eqn:req-domain} and~\eqref{eqn:stakeholders-domain} are the domain equations.

\section{Background} 
\label{sec:Background}

In this section we present all the basic elements required to follow our proposal and the experimental section. We will start with some definitions of the domain of multi-objective optimization followed by the presentation of the classical ILP-based algorithms to find the Pareto front in a bi-objective problem.

\subsection{Multi-objective Optimization}

A multiple criteria optimization problem is defined without loss of generality by
\begin{equation}
\min f(x)=(f_{1}(x),\ldots,f_{p}(x)),\quad \mbox{subject to} \quad x\in X
\end{equation}
where $p\in\mathbb{N},$ $p\geq2$, $f_{i}:X\rightarrow\mathbb{R}$ are the objective functions, $i=1,...,p$, and $X\neq \emptyset$ denotes the feasible solution set. In this article, we consider $X$ discrete and bounded. Every element in $X$ is a vector of dimension $n$, being $n$ the number of variables in the decision space .

The notion of optimality with several objective functions is considered in the sense of Pareto optimization.
A feasible solution $x\in X$ is called \emph{dominated} if there exists another $y\in X$ with $f_{i}(y)\leq f_{i}(x)$ for all $i=1,\ldots,p$, and $f_{k}(y)<f_{k}(x)$ for at least one $k\in\left\{ 1,...,p\right\} $. In this case, $y$ \emph{dominates} $x$ and $x$ \emph{is dominated} by $y$ ($y \preceq x$). If strict inequality holds for all $k\in\left\{ 1,\ldots,p\right\} $, then we say that $x$ is \emph{strictly dominated} by $y$, and $y$ \emph{strictly dominates} $x$ ($y \prec x$) \cite{ehrgott2005multicriteria}.

\theoremstyle{definition}
\begin{definition}
$x\in X$ is an \emph{efficient solution} if there is no $y\in X$ which dominates $x$.
\end{definition}

\begin{definition}
$x\in X$ is called a \emph{weakly efficient solution} if there is no $y\in X$ which strictly dominates $x$.
\end{definition}

The image of an efficient solution $x$, is called a \emph{non-dominated point},  $z=f(x)$. The image of a \emph{weakly efficient solution} $x'$, is called a \emph{weakly non-dominated point}, $z'=f(x')$ .

The set of all efficient solutions of a multiple criteria optimization problem is called \emph{efficient set}, $X_{E}$,  and its image is called Pareto front, $PF=f(X_{E})$. Because many of the elements of $X_{E}$ could lead to the same image, we are only interested in the set $PF$ and one anti-image for every element of this set.

\begin{definition}
An efficient solution is called \emph{supported} if its image lies on the frontier of the convex hull  of $PF\in\mathbb{\mathbb{R}}{}^{p}$. Equivalently, $x\in X$ is supported if it minimizes a weighted sum of the $p$ objectives involving positive weights.
\end{definition}

\begin{definition}
\textbf{(Lexicographic order)} Let $z^{1},z^{2} \in \mathbb{R}^{p}$. We say that $z^{1} <_{lex} z^{2}$ when $z^{1}_{q} < z^{2}_{q}$ for $q = min\{k \mid z^{1}_{k} \neq z^{2}_{k} \}$.
\end{definition}

\begin{definition}
Let $\sigma$ be a permutation, $f:X \rightarrow \mathbb{R}^p$ a vector function, and $f_{\sigma}=\left( f_{\sigma(1)}, f_{\sigma(2)},...,f_{\sigma(p)} \right)$ the vector function based on $f$ where the objectives were re-ordered using permutation $\sigma$. We say that $x \in X$ is a lexicographical optimal solution for permutation $\sigma$ when there is no $y \in X$ with $f_{\sigma}(y) <_{lex} f_{\sigma}(x)$. There exists a maximum of $p!$ lexicographical optimal solutions, one for each permutation.
\end{definition}

The \emph{Next Release Problem} was firstly defined as a bicriteria problem with linear constraints and binary variables in \cite{zhang2007multi}. This is called BOILP (\emph{Bi-Objective Integer Linear Problem}) in the literature. In general, a bicriteria problem has two lexicographical optimal solutions, otherwise it is a trivial problem. Suppose that the images of these two solutions are $z^{1},\,z^{2}\in PF$ with $z_{1}^{1}<z_{1}^{2}$. Then,  $z_{2}^{1}>z_{2}^{2}$, where $z_{1}^{1}=\underset{x\in X}{min}\left\{ f_{1}(x)\right\} $, $z_{2}^{1}=\underset{x\in X}{min}\left\{{ f_{2}(x) \mid f_{1}(x) \leq z_{1}^{1}}\right\}$,
$z_{2}^{2}=\underset{x\in X}{min}\left\{ f_{2}(x)\right\} $ and 
$z_{1}^{2}=\underset{x\in X}{min}\left\{ f_{1}(x) \mid f_{2}(x)\leq z_{2}^{2}\right\}$.

\begin{figure}[!ht]
\begin{center}
\includegraphics[scale=0.55]{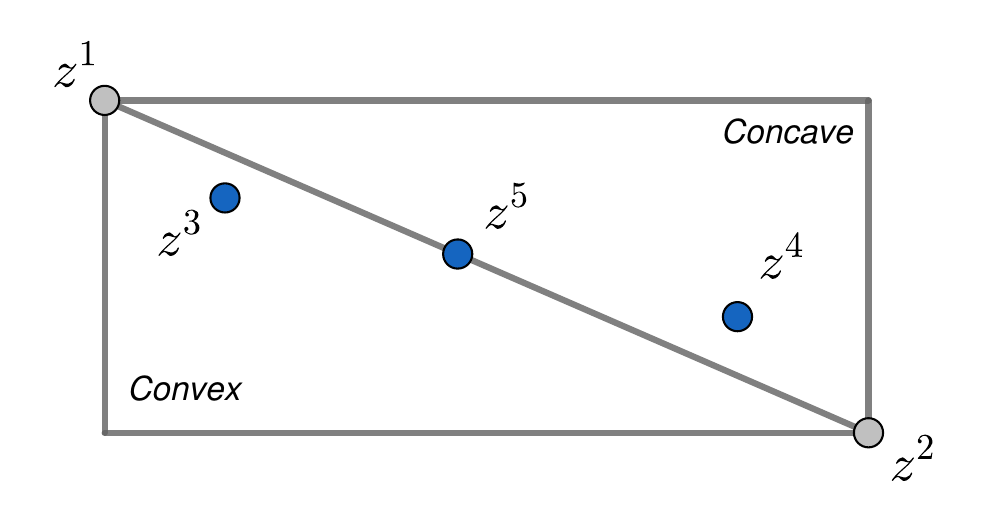} 
\end{center}
\caption{$z^{1}$ and $z^{2}$ form a \emph{box}. The points $z^{3}$ and $z^{5}$ lies in the convex part of the \emph{box}, and the point $z^{4}$ lies in the concave part of the \emph{box}. \label{fig:concave_convex}}
\end{figure}

\begin{definition}
Let $z^{1}$ and $z^{2}$ be two bidimensional points  with $z^{1}_{1} < z^{2}_{1}$. The points $z^1$ and $z^2$ form a box (rectangle) seen in Figure~\ref{fig:concave_convex}. Consider the function 

\begin{equation}
\delta(x)=\lambda_{1}f_{1} (x)+\lambda_{2}f_{2}(x) 
\end{equation}
where $\lambda_{1}=z_{2}^{1}-z_{2}^{2}$ and $\lambda_{2}=z_{1}^{2}-z_{1}^{1}$. Let $y$ be a solution  whose image is inside the box formed by $z^1$ and $z^2$. Then we say that $y$ is in the \emph{convave part} of the box when $\delta(y) >\delta_{0}=\lambda_1 z_1^1 + \lambda_2 z_2^1 = \lambda_1 z_1^2 + \lambda_2 z_2^2$, and is in the \emph{convex part} of the box when $\delta(y)\leq\delta_{0}$.
\end{definition}

\subsection{Classic algorithms for bi-objective optimization} 			
\label{sec:exact_methods}

In this section we present four well-known methods for computing the complete Pareto front of a bi-objective optimization problem. They are the $\varepsilon$-constraint method, the \emph{Augmented $\varepsilon$-constraint} method, \emph{Ehrgott Hybrid's} method and the \emph{Augmented Tchebycheff} method. We also include a description of the dichotomic search used by Veerapen et al.~\cite{veerapen2015integer}, which is the first phase of the Two Phase Method proposed by Ulungu and Teghem~\cite{UlunguTeghem1995}.

\subsubsection{$\varepsilon$-constraint method}
\label{sec:eps}

The general bi-objective $\varepsilon$-constraint method is one of the best-known techniques to solve bicriteria optimization problems. The idea of the algorithm is to minimize one of the objectives while the other is transformed into a constraint~\cite{ehrgott2005multicriteria}. Since there are two objective functions, we can implement two variants of the method, depending on which function we minimize. In general, the result of the method is a set of weakly efficient solutions that must be filtered to find the set of non-dominated points. At the end, this algorithm certifies that the whole Pareto front is found. 
There is another variant of this method which avoids the use of the filtering process. It requires to solve two  subproblems to obtain each efficient solution \cite{berube2009exact}.

\subsubsection{Augmented $\varepsilon$-constraint method}
\label{sec:augmecon}
This method, also called \emph{Augmecon}~\cite{mavrotas2009effective,mavrotas2013improved}, is based on the general $\varepsilon$-constraint method. It adds a new variable and modifies the objective function and one constraint. \emph{Augmecon} is able to obtain one efficient solution with only one call to the underlying single-objective solver. The new created variable has a coefficient $\lambda>0$ in the objective function, and is usually a fixed value in the interval $\left[10^{-6},\,10^{-3}\right]$. For example, if we choose to minimize $f_1$, then the objective function is $f_{1}-\lambda t$ subject to all the constraints of the problem plus $f_{2}(x)+t\leq\varepsilon$. If $\lambda$ is too large, the algorithm could omit solutions. If $\lambda$ is too small, it could generate weakly efficient solutions due to numerical errors in the solver. This value is problem-dependent, but in general, it works well with a value in the range $\left[10^{-6},\,10^{-3}\right]$. 
Compared to $\varepsilon$-constraint, \emph{Augmecon} has the advantages that it ensures that any solution found is efficient and, thus, it requires one single call of the underlying solver per point in the Pareto front.

\subsubsection{Ehrgott's Hybrid method} \label{sec:hybrid}
This method, described in~\cite[p.~101]{ehrgott2005multicriteria}, combines a parameterization of the two objectives with the $\varepsilon$-\emph{constraint} method. It will be called \emph{EHybrid} method. Given a bi-objective optimization problem and two real numbers $\lambda_{1}, \lambda_{2}>0$, at each step the algorithm minimizes $\lambda_{1}f_{1}(x)+\lambda_{2}f_{2}(x)$, subject to the constraints of the problem, $x\in X$, and the new constraints $f_{i}(x)\leq L_{i}$ for $i\in\left\{ 1,2\right\}$, being $L=\left(L_{1},\,L_{2}\right)$ a given point. If the problem has an optimal solution, it must be efficient for the original bi-objective problem.

\begin{figure}[!ht]
\begin{center}
\includegraphics[scale=0.4]{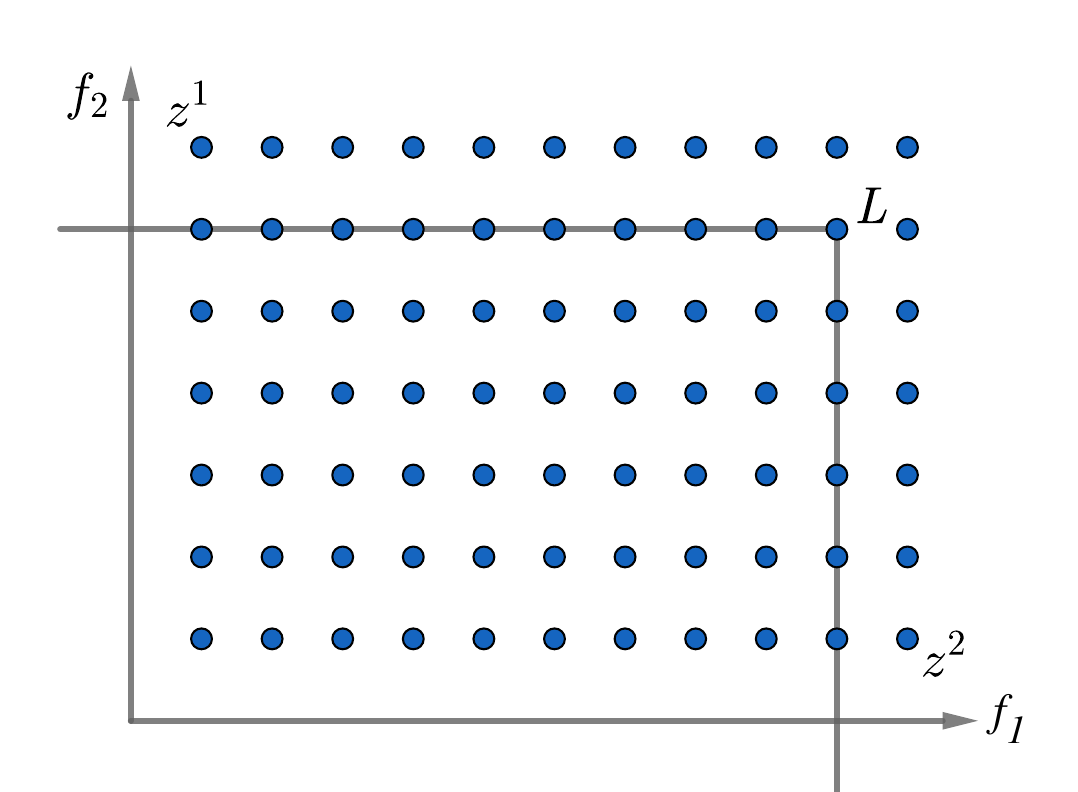}
\par
\end{center}
\caption{For the \emph{EHybrid} method, we analyze a box with two non-dominated points as opposite corners in the current iteration. The vector \emph{L} is selected to look for new non-dominated points inside the box.
\label{fig:hybrid}}
\end{figure}

The \emph{EHybrid} method can start with the lexicographical optimal solutions. At every iteration, it analyzes a box with two adjacent non-dominated points as opposite corners, and looks for a new non-dominated point between them. In Figure~\ref{fig:hybrid}, we can see how the method adds constraints in the two axes when searching for a non-dominated point in a box.
When analyzing the box with corner points $z^{1}$ and $z^{2}$, being $z_{1}^{1}<z_{1}^{2}$, it defines $L=(L_{1},L_{2})$ such that $L_{1}=z_{1}^{2}-\delta$, $L_{2}=z_{2}^{1}-\delta$, where $\delta$ is small enough to avoid omitting solutions. If the subproblem is infeasible, no new non-dominated point exists between them and the box is discarded. If a solution exists, it is efficient for the bi-objective problem and is added to the Pareto Optimal set.

\subsubsection{Augmented Tchebycheff method} 
\label{sec:exact_methods:tchebycheff}

\begin{figure}[!ht]
\begin{center}
\includegraphics[scale=0.4]{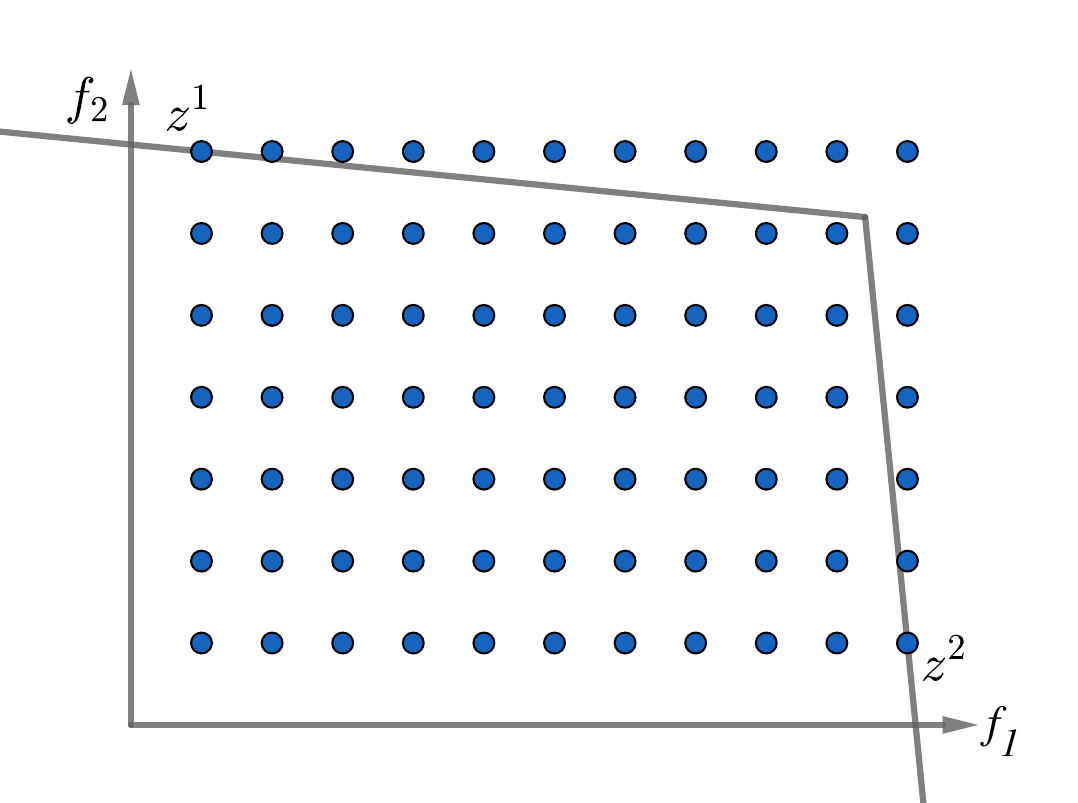} 
\par
\end{center}
\caption{Example of level curve for the \emph{Tchebycheff} method. \label{fig:tchebycheff}}
\end{figure}

This algorithm was introduced by Dächert et al. in \cite{dachert2012augmented} and uses an augmented weighted Tchebycheff norm in order to avoid the generation of weakly non-dominated points. In this paper, this algorithm will be called \emph{Tchebycheff}. 
The method uses as objective function a weighted sum of the $\lVert\cdot\rVert_{\infty}$ metric with an added term using the $\lVert\cdot\rVert_{1}$ metric. This way, we guarantee that a solution to the problem is efficient. The level curves in the objective space for certain value $\alpha$ are unions of linear segments (see Figure~\ref{fig:tchebycheff}).

Starting with the lexicographical optimal solutions, the algorithm analyzes boxes in the objective space, looking for new non-dominated points in between. If the algorithm finds a solution whose image is an extreme of the box it is discarded. Otherwise, the box is broken down into two new boxes to explore. The objective function to minimize is $\max(w_1 z_1(x), w_2 z_2(x)) + \rho \lvert z(x) \rvert$, where $z(x)=f(x)-t$, and $t=\left(t_{1},t_{2}\right)$ is the local ideal point of the box, that is, if we analyze the box with  corners $(z^{1},z^{2})$, then $t=(z_{1}^{1},z_{2}^{2})$. The vector $w$ determines the associated weights of the vector $z$ and the positive real value $\rho$ is fixed and should have the maximum value possible in order to avoid numerical errors in the solver. The values of $w$ and $\rho$ depend on the coordinates of the corners in the current box analyzed.

\subsubsection{Anytime dichotomic search}
\label{sec:exact_methods:ads}

Veerapen et al.~\cite{veerapen2015integer} use an anytime method based on Aneja and Nair's dichotomic scheme~\cite{AnejaNair1979}. The idea is also quite similar to the first phase of the Ulungu and Teghem two-phase method~\cite{UlunguTeghem1995}.  The algorithm starts computing the two lexicographical optimal solutions, form a box with the images of both of them and adds the box to a list of regions to explore. In each iteration of the algorithm the box with the largest diagonal, say $(z^1,z^2)$ is extracted from the list and a weighted sum of the objectives is optimized, where the weights are $\lambda_1=z^1_2-z^2_2$ and $\lambda_2=z^2_1-z^1_1$. With these weights, $z^1$ and $z^2$ evaluate to the same value. If a new solution $z$ is found after this optimization step, its image is added to the Pareto front and is used to divide the box in two new boxes: $(z^1,z)$ and $(z,z^2)$. If no new solution is found the algorithm extracts another box from the list of boxes and iterates again. This process is repeated until a time limit is reached or the list of boxes is empty.

\section{Anytime algorithms}
\label{sec:anytime_methods}

In this section we present the main contribution of this work: five new anytime algorithms based on some of the algorithms described in Section~\ref{sec:exact_methods}. 
The algorithms presented in this section have a similar structure. They start computing the images of the optimal lexicographical solutions and define a set of boxes. Each box is represented by its upper-left and bottom-right corners. 
As long as the set of boxes is not empty, one box is extracted from it, the one with largest area, and it is explored to search for new non-dominated points inside. If a new point is found, two new boxes are to be explored. They  are the result of breaking apart the original box in four pieces and removing the dominated and empty ones. All the algorithms stop when a time limit is reached or when there is no box to explore.

The algorithms of subsections \ref{sec:spf}, \ref{sec:anytchebycheff} and \ref{sec:anyhybrid} can be considered as slight variations of previous algorithms. The algorithms of subsections \ref{sec:anyaugmecon} and \ref{sec:mixed} are completely new, to the authors knowledge.

\subsection{Finding the supported Pareto front} \label{sec:spf}

Our first anytime algorithm, called \emph{SPF} (Supported Pareto Front), focuses the search in the supported efficient solutions. It is a variant of the Anytime Dichotomic Search (\emph{ADS}) of Veerapen et al.~\cite{veerapen2015integer}, which can also find supported efficient solutions only. The main difference between \emph{ADS} and \emph{SPF} is that \emph{ADS} can miss supported efficient points, while \emph{SPF} is designed to find all of them. The code of \emph{SPF} is in Algorithm~\ref{alg:complete_SPF}. In  Line~\ref{lin:spf-opt} we can observe that the scalarized version of the problem contains two constraints for both objective functions. These constraints prevent the algorithm from finding a previously found supported solution, thus, forcing it to find a new one, if it exists. This is the reason why it is able to find all the supported efficient solutions.

Looking at Algorithm \ref{alg:complete_SPF}, after a box is analyzed, if a new solution is found, we check whether its image is in the convex part of the box. If this is the case, we divide the box into two new boxes. If the new solution has its image in the concave part of the box, then it is not supported and is discarded. In Line \ref{alg:SPF:extract} of Algorithm \ref{alg:complete_SPF} we extract the box with the largest area from the set \emph{Boxes}.

\begin{algorithm} [h]
\caption{\emph{SPF}} 	\label{alg:complete_SPF}
\begin{algorithmic} [1]
\STATE $\left\{ z^{1},\,z^{2}\right\} $ $\leftarrow$\ Images of the lexicographical optimal solutions
\STATE $Boxes=\left\{ \left(z^{1},\,z^{2}\right)\right\} $
\STATE \emph{PF =} $\{z^{1}, z^{2}\}$
\WHILE{(\emph{Boxes} $\neq \emptyset$)}
\STATE $\left(\varepsilon^{1},\varepsilon^{2}\right)\leftarrow$ Extract some box from \emph{Boxes}
\label{alg:SPF:extract}
\STATE $\lambda_{1}=\varepsilon_{2}^{1}-\varepsilon_{2}^{2}$ ~~;~~$\lambda_{2}=\varepsilon_{1}^{2}-\varepsilon_{1}^{1}$
\STATE $\left(l_{1},\,l_{2}\right)=\left(\varepsilon_{1}^{2}-1,\,\varepsilon_{2}^{1}-1\right)$

\STATE $P \equiv \min \{ \lambda_{1}f_{1}(x)+\lambda_{2}f_{2}(x) ;\,\textit{s.t.} \, x\in X\,\land\,f_{1}(x)\leq l_{1}$ \, $\land\,f_{2}(x)\leq l_{2} \} $

\label{lin:spf-opt}
\IF{ (\emph{P is feasible})}
\STATE $x^{*}\leftarrow$ Optimal solution of \emph{P}
\STATE $z=\left(f_{1}\left(x^{*}\right),\,f_{2}\left(x^{*}\right)\right)$
\IF {$\left(\lambda_{1}f_{1}(x^{*})+\lambda_{2}f_{2}(x^{*})\leq\lambda_{1}\varepsilon_{1}^{1}+\lambda_{2}\varepsilon_{2}^{2}\right)$} 
\STATE $Boxes=Boxes\,\cup\left\{ \left(z^{1},\,z\right)\right\} \cup\left\{ \left(z,\,z^{2}\right)\right\}$
\STATE \emph{PF = PF $\cup$} $\left\{ z\right\} $
\ENDIF
\ENDIF
\ENDWHILE 
\end{algorithmic}  
\end{algorithm}

\subsection{Anytime version of Augmecon}
\label{sec:anyaugmecon}

The anytime version of Augmecon will be called \emph{AnyAugmecon}\footnote{A preliminary version of this algorithm appeared in the Spanish congress JISBD 2016~\cite{chicano2016estrategias}.} and is presented in Algorithm \ref{alg:anytime_augmecon}. In this and other algorithms we use \emph{obj} to designate the objective we will use as the main objective to optimize in the single-objective formulation of the ILP, and \emph{rest} for the second objective (used as constraint in the ILP formulation). For example, if \emph{obj} $=1$, then \emph{rest} $=2$ and vice versa. We consider \emph{obj} as a parameter of the algorithm.

In addition to the starting objective function and the $\lambda$ value (see Section~\ref{sec:augmecon}) as input parameters, the algorithm fixes $\varepsilon$ as the midpoint between the \emph{rest}-coordinate of the corners of the current box (Line~\ref{alg:anytime_augmecon_epsilon}). Then, it solves the sub-problem. If a new solution is found, it  checks if its image was previously found, which is equivalent to checking if $z^{*}$ is not in the interior region of the box (Line~\ref{alg:anytime_augmecon_not_dominated}). Then, two new boxes are created and added to the set \emph{Boxes}. Otherwise, only one new box is created and added to the set, having half the area of the previous one.

\begin{figure}[!ht]
\begin{center}
\begin{minipage} {0.8\linewidth}
\includegraphics[scale=0.4]{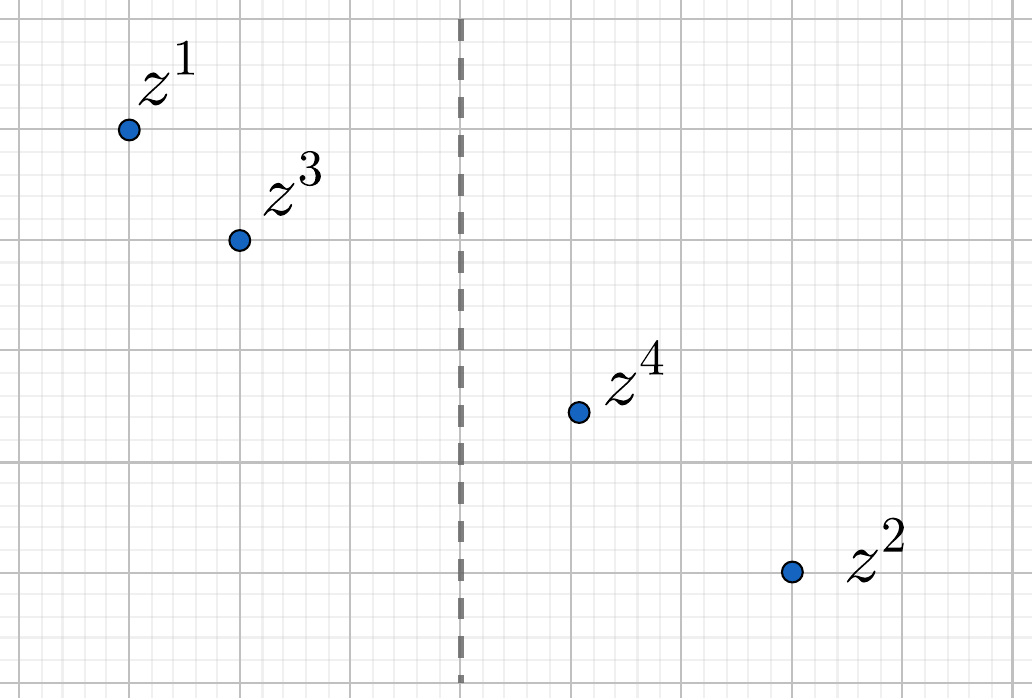} 
\subcaption{\emph{AnyAugmecon $(1, \lambda)$} \label{fig:augmecon1a}}
\end{minipage}
\begin{minipage} {0.8\linewidth}
\includegraphics[scale=0.44]{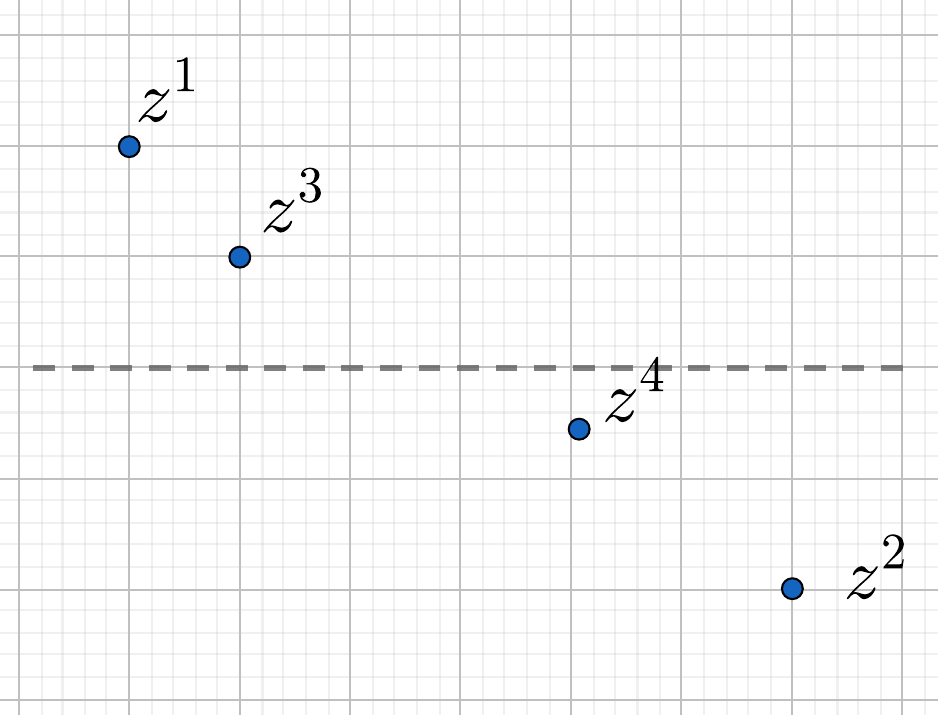}
\subcaption{\emph{AnyAugmecon $(2, \lambda)$} \label{fig:augmecon1b}}
\end{minipage}
\end{center}
\caption{Analyzing a box using \emph{AnyAugmecon $(1,\lambda)$} and \emph{AnyAugmecon $(2,\lambda)$}.}
\label{fig:augmecon1}
\end{figure}

In Figure~\ref{fig:augmecon1} we analyze the box $\left(z^{1},\,z^{2}\right)$ with \emph{AnyAugmecon}. If $f_1$ is optimized ($f_2$ is used as constraint), the algorithm searches from the vertical line to the left, and finds $z^{3}$ as the new point. Then, it analizes \emph{boxes} $\left(z^{1},\,z^{3}\right)$ and $\left(z^{3},\,z^{2}\right)$ (see Figure~\ref{fig:augmecon1a}).
If $f_2$ is optimized, the algorithm searches from the horizontal line to the bottom, obtains $z^{4}$ as the new point and analyzes \emph{boxes} $\left(z^{1},\,z^{4}\right)$ and $\left(z^{4},\,z^{2}\right)$ (see Figure~\ref{fig:augmecon1b}). 

\begin{figure} [!ht]
\begin{center}
\begin{minipage} {0.9\linewidth}
\centering
\includegraphics[scale=0.4]{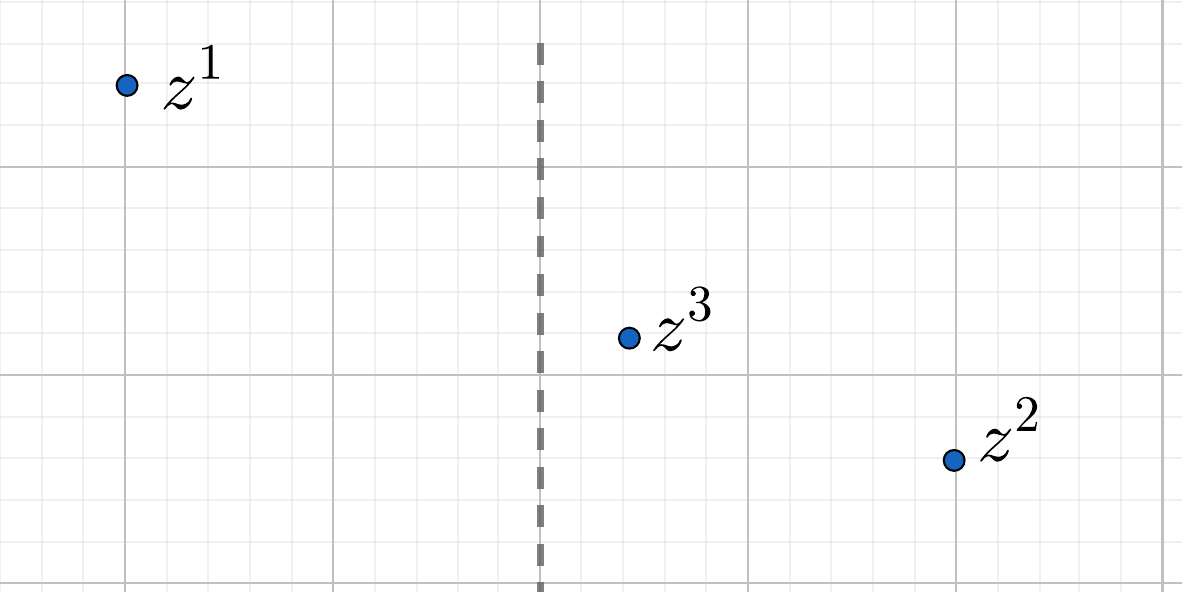} 
\subcaption{No new non-dominated point is found \label{fig:augmecon2a}}
\end{minipage}

\begin{minipage} {0.9\linewidth}
\centering
\includegraphics[scale=0.4]{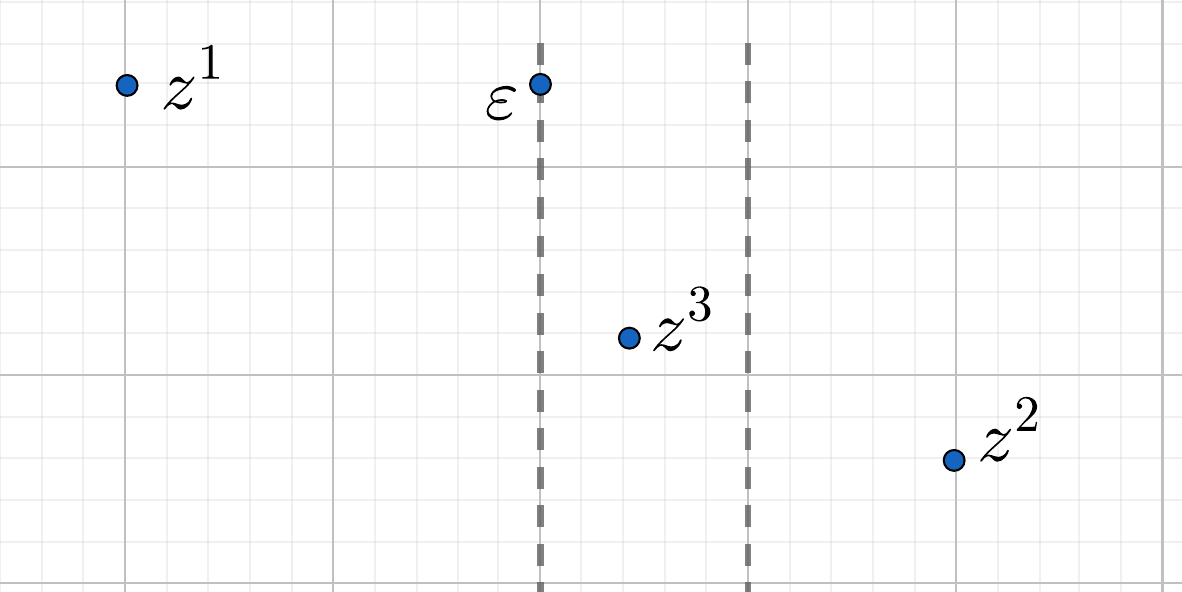}
\subcaption{Then, the algorithm searches in the region determined by the vertical bar ($\varepsilon$) and $z^2$}
\label{fig:augmecon2b}
\end{minipage}
\par
\end{center}
\caption{Use of \emph{AnyAugmecon $(1,\lambda)$} when no new non-dominated point is found.}
 \label{fig:augmecon2}
\end{figure}

In Figure~\ref{fig:augmecon2}, we analyze box $\left(z^{1},\,z^{2}\right)$ with \emph{AnyAugmecon~$(1,\lambda$)}. No new non-dominated point is found, so it explores \emph{box} $\left(\varepsilon,\,z^{2}\right)$ to obtain $z_{3}$, and create the two new \emph{boxes} $\left(\varepsilon,\,z^{3}\right)$ and $\left(z^{3},\,z^{2}\right)$.

\begin{algorithm}[!ht]
\caption{\emph{AnyAugmecon} (\emph{obj},$\lambda)$} 	\label{alg:anytime_augmecon}
\begin{algorithmic} [1]
\STATE $\left\{ z^{1},\,z^{2}\right\} $ $\leftarrow$ Images of lexicographical optimal solutions
\STATE $Boxes=\left\{ \left(z^{1},\,z^{2}\right)\right\} $
\STATE \emph{PF =} $\{z^{1}, z^{2}\}$
\WHILE{$\left( \emph{Boxes} \neq \emptyset \right)$}
\STATE $\left(z^{1},z^{2}\right)\leftarrow$ Extract box with the largest area 
\STATE $\varepsilon=\left(z_{rest}^{1}+z_{rest}^{2}\right)/2$ \label{alg:anytime_augmecon_epsilon}
\STATE $\varepsilon^{*}=z_{rest}^{rest}$ \enskip
\STATE $P \equiv min \left\{ f_{\emph{obj}}(x)-\lambda t \,; \enskip \textit{s.t. } x\in X\land f_{\emph{rest}}(x)+t\leq\varepsilon \right\}$
\IF {  (\emph{P is feasible})}
\STATE $x^{*}\leftarrow$ Solve \emph{P}
\STATE $z^{*}=\left(f_{1}\left(x^{*}\right),\,f_{2}\left(x^{*}\right)\right)$
\IF {$(z^{*} \textit{ is in the interior of the box})$} \label{alg:anytime_augmecon_not_dominated}
\STATE \emph{PF =~PF} \emph{$\cup$} $\left\{ z^{*}\right\} $
\STATE $Boxes=Boxes\,\cup\left\{ \left(z^{1},\,z^{*}\right), \left(z^{*},\,z^{2}\right)\right\} $
\ELSE
\IF {\textbf{ }\emph{(obj = 1)}}
\STATE\textbf{ }$Boxes=Boxes\,\cup\left\{ \left(\left(\varepsilon,z_{1}^{2}\right),\,z^{2}\right)\right\} $
\ELSE \STATE $Boxes=Boxes\,\cup\left\{ \left(z^{1},\left(z_{2}^{1},\varepsilon\right)\right)\right\} $ 
\ENDIF
\ENDIF
\ENDIF
\ENDWHILE 
\end{algorithmic}  
\end{algorithm}

\subsection{Anytime version of Tchebycheff} 
\label{sec:anytchebycheff}

The anytime version of \emph{Tchebycheff}, called \emph{AnyTchebycheff}\footnote{A preliminary version of this algorithm appeared in the Spanish congress JISBD 2016~\cite{chicano2016estrategias}.}, is presented in Algorithm \ref{alg:anytime_tchebycheff}. The only difference with the \emph{Tchebycheff} method is the way it chooses the boxes to be explored next (Line \ref{alg:anytime_tchebycheff_extract}), which is the one with the largest area. This box will be in most of the cases the one increasing the hypervolume ~\cite{fonseca2006improved} in the largest amount.

\begin{algorithm}[!ht]
\caption{\emph{AnyTchebycheff}}
\emph{\label{alg:anytime_tchebycheff}}
\begin{algorithmic}[1]
\STATE $\left\{ z^{1},\,z^{2}\right\} $ $\leftarrow$ Images of lexicographical optimal solutions
\STATE $Boxes=\left\{ \left(z^{1},\,z^{2}\right)\right\} $
\STATE \emph{PF =} $\{z^{1}, z^{2}\}$
\WHILE{(\emph{Boxes} $\neq \emptyset)$ }
\STATE $\left(\varepsilon^{1},\varepsilon^{2}\right)\leftarrow$ Extract box with the largest area \label{alg:anytime_tchebycheff_extract}
\STATE $D=\left\{ x\in \mathbb{R}{}^{2}:x\in X \wedge\,f_{i}(x)\leq\varepsilon_{i}^{3-i},\,i=1,2\right\}$
\STATE $P \equiv min \,\{ \lambda+\rho{\textstyle \sum_{i=1}^{2}\left(f_{i}(x)-t_{i}\right)} \quad \textit{s.t.} \quad x \in D  \quad  \land \qquad $
$ \lambda\geq \underline{w}_{i}(f_{i}(x)-t_{i})\,\,,\,i=1,2 \} $

\IF{ (\emph{P is feasible})}
\STATE $x^{*}\leftarrow$ Optimal solution of \emph{P}
\STATE $z=\left(f_{1}\left(x^{*}\right),\,f_{2}\left(x^{*}\right)\right)$
\STATE \emph{PF = PF $\cup$} $\left\{ z\right\} $
\STATE $Boxes=Boxes\,\cup\left\{ \left(z^{1},\,z\right)\right\} \cup\left\{ \left(z,\,z^{2}\right)\right\}$
\ENDIF
\ENDWHILE 
\end{algorithmic}  
\end{algorithm}

\subsection{Anytime version of the EHybrid method}
\label{sec:anyhybrid}

The anytime variant of the \emph{EHybrid} method commented in Section \ref{sec:hybrid}, called \emph{AnyHybrid}, is presented in Algorithm \ref{alg:anytime_hybrid}. The only difference between them 
is the way in which \emph{AnyHybrid} chooses the boxes to be explored next (Line \ref{alg:anytime_hybrid_extract}), which is the one with the largest area.

\begin{algorithm}[!ht]
\caption{\emph{AnyHybrid}}
\label{alg:anytime_hybrid}
\begin{algorithmic}[1]
\STATE $\left\{ z^{1},\,z^{2}\right\} $ $\leftarrow$ Images of lexicographical optimal solutions
\STATE $Boxes=\left\{ \left(z^{1},\,z^{2}\right)\right\} $
\STATE \emph{PF =} $\{z^{1} ,z^{2}\}$
\WHILE{ (\emph{Boxes} $\neq \emptyset$) }
\STATE $\left(\varepsilon^{1},\varepsilon^{2}\right)\leftarrow$ Extract box with the largest area
\label{alg:anytime_hybrid_extract}
\STATE $\lambda_{1}=\varepsilon_{2}^{1}-\varepsilon_{2}^{2}$ ~~;~~$\lambda_{2}=\varepsilon_{1}^{2}-\varepsilon_{1}^{1}$
\STATE $\left(L_{1},\,L_{2}\right)=\left(\varepsilon_{1}^{2}-1,\,\varepsilon_{2}^{1}-1\right)$

\STATE $P \equiv min \, \{ \lambda_{1}f_{1}(x)+\lambda_{2}f_{2}(x);$
 \quad $\emph{s.t. } \quad x\in X \quad \land\ \qquad $
 $f_{i}(x)\leq L_{i}\,\,,\,i=1,2 \} $

\IF{ (\emph{P is feasible}) }
\STATE $x^{*}\leftarrow$ Optimal solution of \emph{P}
\STATE $z=\left(f_{1}\left(x^{*}\right),\,f_{2}\left(x^{*}\right)\right)$
\STATE \emph{PF = PF $\cup$} $\left\{ z\right\} $
\STATE $Boxes=Boxes\,\cup\left\{ \left(z^{1},\,z\right)\right\} \cup\left\{ \left(z,
\,z^{2}\right)\right\} $
\ENDIF
\ENDWHILE 
\end{algorithmic}  
\end{algorithm}

Recall that in the \emph{AnyHybrid} algorithm, an explored box could contain no non-dominated points, in which case, the box is discarded. In \emph{AnyAugmecon} and \emph{AnyTchebycheff} a solution is always found, maybe a new one with image in the interior of the box, or maybe a repeated or dominated solution previously found. In all the algorithms that we have exposed so far, when a new non-dominated point is found, two new boxes are generated, but in \emph{AnyAugmecon}, if the image of the new solution is not inside the box, only one reduced box is created.

\subsection{Mixed anytime algorithm}
\label{sec:mixed}

We present in this section two variants of an algorithm that combines two anytime methods on-the-fly: \emph{AnyHybrid} and \emph{AnyTchebycheff}. The selection on which approach to use depends on the solution found in the previous iteration. \emph{EHybrid} method is fast but not very good for concave fronts, and \emph{Tchebycheff} method is good finding spread solutions, so it seems natural to combine their anytime variants together in one algorithm. Before presenting the algorithm, we need to introduce a definition.

\begin{definition}
Let $a$ and $b$ two integer numbers, with $a<b$. Let $c$ be such that $a<c<b$. We say that $c$ is \emph{close to} $a$ if $c-a<\frac{1}{4}(b-a)$. 
We say that $c$ is \emph{close to} $b$ if $b-c<\frac{1}{4}(b-a)$. 
\end{definition}

\begin{figure}[!ht]
\begin{centering}
\includegraphics[scale=1]{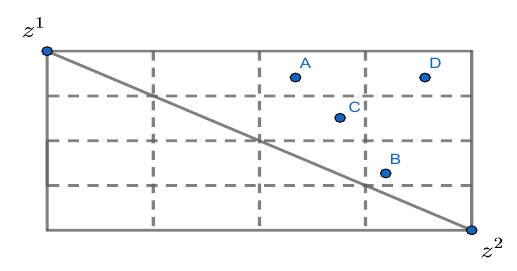}
\par
\end{centering}
\caption{For \emph{MixHT}, if the current point is in the \emph{convex} part of the \emph{box}, the two new \emph{boxes} will use \emph{AnyHybrid} in the next iteration. For simplicity, name \textbf{H} to \emph{AnyHybrid} and \textbf{T} to \emph{AnyTchebycheff}. 
If the new point is $A$, \emph{box} $(z^{1},A)$ will use \textbf{H} and \emph{box} $(A,z^{2})$ will use \textbf{T}.
If the new point is $B$, \emph{box} $(z^{1},B)$ will use \textbf{T} and \emph{box} $(B,z^{2})$ will use \textbf{H}.
If the new point is $C$, \emph{boxes} $(z^{1},C)$ and $(C,z^{2})$ will use \textbf{H}.
If the new point is point $D$, \emph{boxes} $(z^{1},D)$ and $(D,z^{2})$ will use \textbf{T}.}
\label{fig:mixed}
\end{figure}

Now we apply this definition to our problem. Depending on how close is every component of the image in the new solution with respect to the corners of the box, we choose one algorithm or the other for the next iteration. Suppose that the box is $(z^{1},z^{2})$ and the new point $z=(z_{1},z_{2})$ is in the \emph{concave} part. If $z_{1}$ is \emph{close to} $z_{1}^{1}$ or \emph{close to}  $z_{1}^{2}$, the box $(z^{1},z)$ will be analyzed using \emph{AnyTchebycheff}, otherwise \emph{AnyHybrid} will be used. Regarding box $(z,z^{2})$, we check if $z_{2}$ is \emph{close to} $z_{2}^{1}$ or $z_{2}^{2}$. If this is the case, it will be analyzed using \emph{AnyTchebycheff}, otherwise \emph{AnyHybrid} is used. See Figure~\ref{fig:mixed} for a graphical example. 
This new algorithm will be called \emph{MixHT} and its pseudocode is shown in Algorithm \ref{alg:mixed_1}. In Algorithm \emph{MixHT} we associate to the boxes the algorithm (\emph{AnyHybrid} or \emph{AnyTchebycheff}) to be used in its exploration (see Line 4 of Algorithm~\ref{alg:mixed_1}).

\begin{algorithm}[!ht]
\caption{\emph{MixHT}}
\label{alg:mixed_1}
\begin{algorithmic}[1]
\STATE $\left\{ z^{1},\,z^{2}\right\} $ $\leftarrow$ Images of lexicographical optimal solutions
\STATE \emph{H} $\leftarrow$ \emph{AnyHybrid} method
\STATE \emph{T} $\leftarrow$ \emph{AnyTchebycheff} method
\STATE $Boxes=\left\{ \left(z^{1},\,z^{2},\,H\right)\right\}$ 
\STATE \emph{PF =} $\{z^{1}, z^{2}\}$
\WHILE{ ($Boxes\neq \emptyset$)}
\STATE $\left(z^{1},\,z^{2},\,alg\right)\leftarrow$ Extract box with the largest area
\STATE Explore box $(z^1, z^2)$ using \emph{alg} method
\IF{ (\emph{Problem is feasible})}
\STATE $x^{*}\leftarrow$ Optimal solution
\STATE $z=\left(f_{1}\left(x^{*}\right),\,f_{2}\left(x^{*}\right)\right)$
\STATE \emph{PF = PF $\cup$} $\left\{ z\right\}$
\IF {$\left(z_{1} \textit{ is close to } z_{1}^{2} \right)$ \textbf{and} ($z_{1}$ is in the \emph{concave} part)}
\STATE $B_{1}=\left(z^{1},\,z,\,T \right)$
\ELSE 
\STATE $B_{1}=\left(z^{1},\,z,\,H \right)$
\ENDIF

\IF {$\left(z_{2} \textit{ is close to } z_{2}^{1} \right)$ \textbf{and} ($z_{2}$ is in the \emph{concave} part)}
\STATE $B_{2}=\left(z,\,z^{2},\,T \right)$
\ELSE
\STATE $B_{2}=\left(z,\,z^{2},\,H \right)$
\ENDIF
\STATE $Boxes=Boxes\,\cup B_{1} \cup B_{2}$
\ENDIF
\ENDWHILE 
\end{algorithmic}  
\end{algorithm}

The second variant presented in this section, called \emph{MixSHT}, is shown in Algorithm~\ref{alg:mixed_2} and is similar to the previous one, with the difference that the supported non-dominated points are obtained first. The rationale behind this approach is that the supported solutions give a very good hypervolume. After that, we analyze the non-supported points using \emph{MixHT} algorithm. We begin exploring the entire supported Pareto front using \emph{SPF}, and every time we find a non-supported point, we store it in another set, which will be processed later. Therefore, we need to consider two sets of boxes, named $Boxes_{1}$ and $Boxes_{2}$ in Algorithm~\ref{alg:mixed_2}. The images of the lexicographical optimal solutions are stored in $Boxes_{1}$. $Boxes_{2}$  will be explored after $Boxes_{1}$ is empty. When analyzing a \emph{box} of $Boxes_{1}$, if the image of a solution is in the convex part, the two new \emph{boxes} are added to  $Boxes_{1}$. In this way, we are first getting all the supported solutions. However, if we find a solution with image in the concave part of the box, the two new boxes to be explored are stored in $Boxes_{2}$, which will be explored after $Boxes_{1}$ has been exhausted. The exploration of  $Boxes_{2}$ is done using \emph{MixHT} algorithm.

\begin{algorithm}
\caption{\emph{MixSHT}}
\label{alg:mixed_2}
\begin{algorithmic}[1]
\STATE $\left\{ z^{1},\,z^{2}\right\} $ $\leftarrow$ Images of lexicographical optimal solutions
\STATE \emph{H} $\leftarrow$ \emph{AnyHybrid} method; \, \emph{T} $\leftarrow$ \emph{AnyTchebycheff} method

\STATE $Boxes_{1}=\left\{ \left(z^{1},\,z^{2},\,H\right)\right\}; \quad Boxes_{2}= \emptyset$ 
\STATE \emph{PF =} $\{z^{1},z^{2}\}$
\WHILE{ $\left(Boxes_{1} \neq \emptyset \right)$}

\STATE $\left(z^{1},\,z^{2},\,alg\right)\leftarrow$ Extract a box with the largest area from $Boxes_{1}$ 
\STATE Explore box $(z^1,z^2)$ using \emph{alg} method
\IF{ (\emph{Problem is feasible})}
\STATE $x^{*}\leftarrow$ Optimal solution ;\, $z=\left(f_{1}\left(x^{*}\right),\,f_{2}\left(x^{*}\right)\right)$

\STATE \emph{PF = PF $\cup$} $\left\{ z\right\}$
\IF {(z is in the \emph{convex} part)}
\STATE $Boxes_{1}=Boxes_{1}\,\cup (z^{1},\,z,\,H) \cup (z,\,z^{2},\,H)$
\ELSE
\STATE $Boxes_{2}=Boxes_{2}\,\cup (z^{1},\,z,\,H) \cup (z,\,z^{2},\,H)$
\ENDIF
\ENDIF
\ENDWHILE 

\WHILE{ $\left(Boxes_{2} \neq \emptyset \right)$}

\STATE $\left(z^{1},\,z^{2},\,alg\right)\leftarrow$ Extract a box with the largest area from $Boxes_{2}$ 
\STATE Explore box $(z^1,z^2)$ using \emph{alg} method
\IF{ (\emph{Problem is feasible})}
\STATE $x^{*}\leftarrow$ Optimal solution; \, $z=\left(f_{1}\left(x^{*}\right),\,f_{2}\left(x^{*}\right)\right)$

\STATE \emph{PF = PF $\cup$} $\left\{ z\right\}$

\STATE \textbf{if} $\left(z_{1} \textit{ is close to } z_{1}^{2} \right)$ \textbf{and} ($z_{1}$ is in the \emph{concave} part) \textbf{then} \,
$B_{1}=\left(z^{1},\,z,\,T \right)$
\STATE \textbf{else} \, $B_{1}=\left(z^{1},\,z,\,H \right)$

\STATE \textbf{if} $\left(z_{2} \textit{ is close to } z_{2}^{1} \right)$ \textbf{and} ($z_{2}$ is in the \emph{concave} part) \textbf{then} \,
$B_{2}=\left(z,\,z^{2},\,T \right)$
\STATE \textbf{else} \, $B_{2}=\left(z,\,z^{2},\,H \right)$

\STATE $Boxes_{2}=Boxes_{2}\,\cup B_{1} \cup B_{2}$
\ENDIF
\ENDWHILE
\end{algorithmic}  
\end{algorithm}

\section{Analysis and computational results}
\label{sec:computational_results}

In order to answer our research questions we perform a thorough experimental study using well-known benchmarks for Next Release Problems and the algorithms defined in the previous sections.

\subsection{Instances and parameters}

To perform the computational experiments, we used the instances presented in \cite{veerapen2015integer} which were previously described in \cite{xuan2012solving}. They are divided into two groups of datasets, called \emph{classic instances} and \emph{realistic instances}. The first group is composed of five synthetic datasets named \emph{nrp1} to \emph{nrp5}. The \emph{realistic instances} use the bug repositories for the Eclipse, Gnome, and Mozilla open-source projects. Four subsets of bugs were extracted from the three repositories (\emph{nrp-e1} to \emph{nrp-e4}, \emph{nrp-g1} to \emph{nrp-g4}, \emph{nrp-m1} to \emph{nrp-m4}). The requirements for \emph{realistic instances} do not have prerequisites. The number of requirements and the stakeholders for every dataset is shown in Table~{\ref{tab:requerimentsandstakeholders}.

\begin{table}%[!hb]
\centering
\begin{tabular}{|c|rr||c|rr|}
\hline 
Dataset & \emph{req}  & \emph{stake} & Dataset & \emph{req} & \emph{stake} \tabularnewline
\hline 
\emph{nrp1} & \numprint{140} & \numprint{100} & \emph{nrp-g1} & \numprint{2690} & \numprint{445}\tabularnewline
\emph{nrp2} & \numprint{620} & \numprint{500} & \emph{nrp-g2} & \numprint{2650} & \numprint{315}\tabularnewline
\emph{nrp3} & \numprint{1500} & \numprint{500} & \emph{nrp-g3} & \numprint{2512} & \numprint{423}\tabularnewline
\emph{nrp4} & \numprint{3250} & \numprint{750} & \emph{nrp-g4} & \numprint{2246} & \numprint{294}\tabularnewline
\emph{nrp5} & \numprint{1500} & \numprint{1000} & \emph{nrp-m1} & \numprint{4060} & \numprint{768}\tabularnewline
\emph{nrp-e1} & \numprint{3502} & \numprint{536} & \emph{nrp-m2} & \numprint{4368} & \numprint{617}\tabularnewline
\emph{nrp-e2} & \numprint{4254} & \numprint{491} & \emph{nrp-m3} & \numprint{3566} & \numprint{765}\tabularnewline
\emph{nrp-e3} & \numprint{2844} & \numprint{456} & \emph{nrp-m4} & \numprint{3643} & \numprint{568}\tabularnewline
\emph{nrp-e4} & \numprint{3186} & \numprint{399} &  &  & \tabularnewline
\hline 
\end{tabular}
\caption{Number of requirements and stakeholders for every dataset \label{tab:requerimentsandstakeholders}}
\end{table}

We use the hypervolume indicator \cite{fonseca2006improved} as a measure of the objective space that is dominated by the points computed by the algorithms. In the bi-objective case, it represents the union of the regions of all the rectangles that are dominated by the non-dominated points. As a reference point to calculate the hypervolume we consider the Nadir point $(z_{1}^{2},z_{2}^{1})$, being $z^{1}$ and $z^{2}$ the images of the lexicographical optimal solutions.
To solve the NRP instances we use a 1~GHz CPU machine, with four cores and 16 GB of RAM. We programmed the algorithms using C++ and CPLEX 12.6.2. Source code is available at https://github.com/MiguelAngelDominguezRios/anytime-nrp. We set the CPLEX parameters, \begin{footnotesize} CPXPARAMEPGAP $=$ CPXPARAMEPAGAP $=$ CPXPARAMEPINT $= 0$ \end{footnotesize}, as used in \cite{veerapen2015integer}. All the anytime methods are stopped after 60 seconds of computation, unless a different stopping condition is indicated.

Although all the algorithms are deterministic, the number of solutions found can differ for different executions, also for a fixed time. In each call to the solver, CPLEX manages some internal parameters, such as the remaining available memory of the machine, which can have an influence in the tree exploration to obtain the next solution. This explains why we can obtain a different number of solutions even when we execute the same instance for the same amount of time.

To check these variations, we did 30 executions for every instance and for every algorithm, and then use the average values. The \emph{Pearson} coefficient, $\sigma /\mu$ (standard deviation divided by the average), does not exceed the amount of 2\% in the worst case, so the results can be considered stable. If a solution is found after the runtime limit (60 seconds), it is discarded. 

\subsection{Answering RQ1: Comparison of anytime methods}
 \label{sec:computational_anytime}

In Table~\ref{tab:spf}, we execute the \emph{SPF} algorithm without time limit to calculate the complete supported Pareto front for all NRP instances. We indicate the number of supported non-dominated points found by Algorithm \ref{alg:complete_SPF}, which is exact, and by \cite{veerapen2015integer}, which is approximate. The reader can observe the great difference of the total percentage of solutions found in every algorithm. The largest difference occurs in \emph{nrp5}, where \emph{SPF} finds around three times the number of supported solutions of \emph{ADS}. We can also observe that the number of supported solutions can be as low as 2\% of the total number of non-dominated solutions.

\begin{table}[!hb]

	\centering
\begin{tabular}{|c|r|r|r|r|r|}
\hline 
Dataset & $\lvert \emph{PF} \rvert$ & $\lvert \emph{SPF} \rvert$ & $\lvert \emph{ADS} \rvert$ & $\frac{\lvert \emph{SPF} \rvert}{\lvert \emph{PF} \rvert} \%$ & $\frac{\lvert \emph{ADS} \rvert}{\lvert \emph{PF} \rvert} \%$ \tabularnewline
\hline 
\emph{nrp1} & \numprint{465} & 28 & 27 & 6.0 & 5.8\tabularnewline
\emph{nrp2} & \numprint{4540} & 89 & 70 & 2.0 & 1.5\tabularnewline
\emph{nrp3} & \numprint{6296} & 246 & 172 & 3.9 & 2.7\tabularnewline
\emph{nrp4} & \numprint{13489} & 276 & 195 & 2.0 & 1.5\tabularnewline
\emph{nrp5} & \numprint{2898} & 781 & 264 & 27.0 & 9.1\tabularnewline
\emph{nrp-e1} & \numprint{10331} & 826 & 309 & 8.0 & 3.0\tabularnewline
\emph{nrp-e2} & \numprint{10573} & 680 & 300 & 6.4 & 2.8\tabularnewline
\emph{nrp-e3} & \numprint{8344} & 600 & 268 & 7.2 & 3.2\tabularnewline
\emph{nrp-e4} & \numprint{8303} & 454 & 257 & 5.5 & 3.1\tabularnewline
\emph{nrp-g1} & \numprint{9280} & 778 & 233 & 8.4 & 2.5\tabularnewline
\emph{nrp-g2} & \numprint{6393} & 341 & 209 & 5.3 & 3.3\tabularnewline
\emph{nrp-g3} & \numprint{8457} & 603 & 228 & 7.1 & 2.7\tabularnewline
\emph{nrp-g4} & \numprint{6171} & 544 & 201 & 8.8 & 3.3\tabularnewline
\emph{nrp-m1} & \numprint{13773} & \numprint{1252} & 351 & 9.1 & 2.6\tabularnewline
\emph{nrp-m2} & \numprint{12933} & 760 & 329 & 5.9 & 2.5\tabularnewline
\emph{nrp-m3} & \numprint{12624} & \numprint{1059} & 324 & 8.4 & 2.6\tabularnewline
\emph{nrp-m4} & \numprint{11547} & 995 & 295 & 8.6 & 2.6\tabularnewline
\hline 
\end{tabular}
\caption{Comparing the number of supported solutions obtained in NRP instances using  \emph{SPF} and \emph{ADS}. \label{tab:spf}}
\end{table}

Now we study the results of the remaining four anytime algorithms for the NRP instances. We use two variants for \emph{AnyAugmecom}, each one with a different objective function as main goal to optimize in the subproblem and the parameter $\lambda$ as used in \cite{chicano2016estrategias}. We also use the two variants described for the Mixed algorithm combining \emph{EHybrid} and \emph{Tchebycheff}. In total, we compare six algorithms  in our experiments.
As a quality of the solution to measure, we consider the percentage of the total hypervolume in the criteria space and the percentage of total solutions. These results are displayed in Table~\ref{tab:anytime_60}.

% Table generated by ExcelLaTeX from sheet 'tabla_anytime_60'
\begin{table*}

\small
  \centering
  \begin{tabular}{|c|c|rrrrrr|}
\cline{3-8}    
\multicolumn{1}{c}{} & \multicolumn{1}{c|}{ } & 
\multicolumn{1}{c|}{\begin{sideways}\textit{AnyAugmecon (1,$\lambda$)\ \ }\end{sideways}} & 
\multicolumn{1}{c|}{\begin{sideways}\textit{AnyAugmecon (2,$\lambda$) }\end{sideways}} & 
\multicolumn{1}{c|}{\begin{sideways}\textit{AnyHybrid}\end{sideways}} & 
\multicolumn{1}{c|}{\begin{sideways}\textit{AnyTchebycheff}\end{sideways}} & 
\multicolumn{1}{c|}{\begin{sideways}\textit{MixHT}\end{sideways}} & 
\multicolumn{1}{c|}{\begin{sideways}\textit{MixSHT}\end{sideways}} \\
\hhline{--------}

\multirow{2}{*}{\emph{nrp1}} & \%Hyper  & \cellcolor[rgb]{ .847,  .847,  .847}\textbf{100.000} & \cellcolor[rgb]{ .847,  .847,  .847}\textbf{100.000} & \cellcolor[rgb]{ .847,  .847,  .847}\textbf{100.000} & \cellcolor[rgb]{ .847,  .847,  .847}\textbf{100.000} & \cellcolor[rgb]{ .847,  .847,  .847}\textbf{100.000} & \cellcolor[rgb]{ .847,  .847,  .847}\textbf{100.000} \\
\hhline{|~|~|-|-|-|-|-|-|}
          & \%PF  & \textbf{100.000} & \textbf{100.000} & \textbf{100.000} & \textbf{100.000} & \textbf{100.000} & \textbf{100.000} \\
\hhline{--------}

\multirow{2}{*}{\emph{nrp2}} & \%Hyper  & \cellcolor[rgb]{ .847,  .847,  .847}97.673 & \cellcolor[rgb]{ .847,  .847,  .847}\textbf{98.869} & \cellcolor[rgb]{ .847,  .847,  .847}60.202 & \cellcolor[rgb]{ .847,  .847,  .847}97.107 & \cellcolor[rgb]{ .847,  .847,  .847}96.403 & \cellcolor[rgb]{ .847,  .847,  .847}96.667 \\
\hhline{|~|~|-|-|-|-|-|-|}
          & \%PF  & 1.0   & 2.0   & 3.6   & 0.7   & 1.1   & \textbf{4.6} \\
\hhline{--------}

\multirow{2}{*}{\emph{nrp3}} & \%Hyper  & \cellcolor[rgb]{ .847,  .847,  .847}99.714 & \cellcolor[rgb]{ .847,  .847,  .847}99.689 & \cellcolor[rgb]{ .847,  .847,  .847}99.694 & \cellcolor[rgb]{ .847,  .847,  .847}99.375 & \cellcolor[rgb]{ .847,  .847,  .847}\textbf{99.740} & \cellcolor[rgb]{ .847,  .847,  .847}99.557 \\
\hhline{|~|~|-|-|-|-|-|-|}
          & \%PF  & 4.5   & 4.1   & 5.6   & 2.1   & 5.4   & \textbf{8.3} \\
\hhline{--------}

\multirow{2}{*}{\emph{nrp4}} & \%Hyper  & \cellcolor[rgb]{ .847,  .847,  .847}\textbf{98.862} & \cellcolor[rgb]{ .847,  .847,  .847}97.847 & \cellcolor[rgb]{ .847,  .847,  .847}90.546 & \cellcolor[rgb]{ .847,  .847,  .847}94.586 & \cellcolor[rgb]{ .847,  .847,  .847}97.464 & \cellcolor[rgb]{ .847,  .847,  .847}90.396 \\
\hhline{|~|~|-|-|-|-|-|-|}
          & \%PF  & 0.6   & 0.3   & 0.6   & 0.1   & 0.4   & \textbf{2.0} \\
\hhline{--------}

\multirow{2}{*}{\emph{nrp5}} & \%Hyper  & \cellcolor[rgb]{ .847,  .847,  .847} {*} & \cellcolor[rgb]{ .847,  .847,  .847}\textbf{99.969} & \cellcolor[rgb]{ .847,  .847,  .847}99.953 & \cellcolor[rgb]{ .847,  .847,  .847}67.838 & \cellcolor[rgb]{ .847,  .847,  .847}99.822 & \cellcolor[rgb]{ .847,  .847,  .847}99.873 \\
\hhline{|~|~|-|-|-|-|-|-|}
          & \%PF  &  *     & \textbf{45.5} & 36.8  & 0.1   & 14.6  & 39.6 \\
\hhline{--------}
 
\multirow{2}{*}{\emph{nrp-e1}} & \%Hyper  & \cellcolor[rgb]{ .847,  .847,  .847}99.896 & \cellcolor[rgb]{ .847,  .847,  .847}99.872 & \cellcolor[rgb]{ .847,  .847,  .847}\textbf{99.898} & \cellcolor[rgb]{ .847,  .847,  .847}99.737 & \cellcolor[rgb]{ .847,  .847,  .847}99.897 & \cellcolor[rgb]{ .847,  .847,  .847}99.882 \\
\hhline{|~|~|-|-|-|-|-|-|}
          & \%PF  & 6.5   & 5.4   & 7.9   & 2.7   & 7.1   & \textbf{8.5} \\
\hhline{--------}
 
\multirow{2}{*}{\emph{nrp-e2}} & \%Hyper  & \cellcolor[rgb]{ .847,  .847,  .847}99.860 & \cellcolor[rgb]{ .847,  .847,  .847}99.758 & \cellcolor[rgb]{ .847,  .847,  .847}\textbf{99.882} & \cellcolor[rgb]{ .847,  .847,  .847}99.625 & \cellcolor[rgb]{ .847,  .847,  .847}99.877 & \cellcolor[rgb]{ .847,  .847,  .847}99.855 \\
\hhline{|~|~|-|-|-|-|-|-|}
          & \%PF  & 4.7   & 2.8   & 5.5   & 2.0   & 5.2   & \textbf{5.7} \\
\hhline{--------}

\multirow{2}{*}{\emph{nrp-e3}} & \%Hyper  & \cellcolor[rgb]{ .847,  .847,  .847}\textbf{99.931} & \cellcolor[rgb]{ .847,  .847,  .847}99.925 & \cellcolor[rgb]{ .847,  .847,  .847}99.922 & \cellcolor[rgb]{ .847,  .847,  .847}99.831 & \cellcolor[rgb]{ .847,  .847,  .847}99.920 & \cellcolor[rgb]{ .847,  .847,  .847}99.896 \\
\hhline{|~|~|-|-|-|-|-|-|}
          & \%PF  & 11.5  & 10.6  & \textbf{13.8} & 5.3   & 11.3  & 13.0 \\
\hhline{--------}
    
\multirow{2}{*}{\emph{nrp-e4}} & \%Hyper  & \cellcolor[rgb]{ .847,  .847,  .847}\textbf{99.911} & \cellcolor[rgb]{ .847,  .847,  .847}99.860 & \cellcolor[rgb]{ .847,  .847,  .847}99.900 & \cellcolor[rgb]{ .847,  .847,  .847}99.773 & \cellcolor[rgb]{ .847,  .847,  .847}99.895 & \cellcolor[rgb]{ .847,  .847,  .847}99.878 \\
\hhline{|~|~|-|-|-|-|-|-|}
          & \%PF  & 9.2   & 6.1   & \textbf{10.6} & 4.1   & 8.8   & 10.4 \\
\hhline{--------}
    
\multirow{2}{*}{\emph{nrp-g1}} & \%Hyper  & \cellcolor[rgb]{ .847,  .847,  .847}\textbf{99.955} & \cellcolor[rgb]{ .847,  .847,  .847}99.945 & \cellcolor[rgb]{ .847,  .847,  .847}99.948 & \cellcolor[rgb]{ .847,  .847,  .847}99.896 & \cellcolor[rgb]{ .847,  .847,  .847}99.946 & \cellcolor[rgb]{ .847,  .847,  .847}99.925 \\
\hhline{|~|~|-|-|-|-|-|-|}
          & \%PF  & 15.4  & 13.1  & \textbf{18.2} & 7.8   & 14.5  & 15.8 \\
\hhline{--------}

\multirow{2}{*}{\emph{nrp-g2}} & \%Hyper  & \cellcolor[rgb]{ .847,  .847,  .847}\textbf{99.956} & \cellcolor[rgb]{ .847,  .847,  .847}99.934 & \cellcolor[rgb]{ .847,  .847,  .847}99.951 & \cellcolor[rgb]{ .847,  .847,  .847}99.892 & \cellcolor[rgb]{ .847,  .847,  .847}99.945 & \cellcolor[rgb]{ .847,  .847,  .847}99.941 \\
\hhline{|~|~|-|-|-|-|-|-|}
          & \%PF  & 19.2  & 14.0  & \textbf{22.7} & 9.6   & 17.5  & 17.3 \\
\hhline{--------}
 
\multirow{2}{*}{\emph{nrp-g3}} & \%Hyper  & \cellcolor[rgb]{ .847,  .847,  .847}\textbf{99.963} & \cellcolor[rgb]{ .847,  .847,  .847}99.955 & \cellcolor[rgb]{ .847,  .847,  .847}99.955 & \cellcolor[rgb]{ .847,  .847,  .847}99.912 & \cellcolor[rgb]{ .847,  .847,  .847}99.954 & \cellcolor[rgb]{ .847,  .847,  .847}99.943 \\
\hhline{|~|~|-|-|-|-|-|-|}
          & \%PF  & 19.1  & 16.2  & \textbf{22.1} & 9.7   & 17.3  & 17.4 \\
\hhline{--------}

\multirow{2}{*}{\emph{nrp-g4}} & \%Hyper  & \cellcolor[rgb]{ .847,  .847,  .847}\textbf{99.969} & \cellcolor[rgb]{ .847,  .847,  .847}99.956 & \cellcolor[rgb]{ .847,  .847,  .847}99.957 & \cellcolor[rgb]{ .847,  .847,  .847}99.924 & \cellcolor[rgb]{ .847,  .847,  .847}99.960 & \cellcolor[rgb]{ .847,  .847,  .847}99.948 \\
\hhline{|~|~|-|-|-|-|-|-|}
          & \%PF  & 27.0  & 20.8  & \textbf{28.9} & 14.0  & 23.6  & 23.2 \\
\hhline{--------}

\multirow{2}{*}{\emph{nrp-m1}} & \%Hyper  & \cellcolor[rgb]{ .847,  .847,  .847}99.802 & \cellcolor[rgb]{ .847,  .847,  .847}99.792 & \cellcolor[rgb]{ .847,  .847,  .847}99.794 & \cellcolor[rgb]{ .847,  .847,  .847}99.449 & \cellcolor[rgb]{ .847,  .847,  .847}\textbf{99.809} & \cellcolor[rgb]{ .847,  .847,  .847}99.791 \\
\hhline{|~|~|-|-|-|-|-|-|}
          & \%PF  & 2.8   & 2.7   & 3.4   & 1.0   & 3.2   & \textbf{3.7} \\
\hhline{--------}

\multirow{2}{*}{\emph{nrp-m2}} & \%Hyper  & \cellcolor[rgb]{ .847,  .847,  .847}99.792 & \cellcolor[rgb]{ .847,  .847,  .847}99.721 & \cellcolor[rgb]{ .847,  .847,  .847}99.816 & \cellcolor[rgb]{ .847,  .847,  .847}99.442 & \cellcolor[rgb]{ .847,  .847,  .847}99.825 & \cellcolor[rgb]{ .847,  .847,  .847}\textbf{99.825} \\
\hhline{|~|~|-|-|-|-|-|-|}
          & \%PF  & 2.8   & 2.1   & 3.5   & 1.1   & 3.4   & \textbf{3.7} \\
\hhline{--------}

\multirow{2}{*}{\emph{nrp-m3}} & \%Hyper  & \cellcolor[rgb]{ .847,  .847,  .847}99.815 & \cellcolor[rgb]{ .847,  .847,  .847}\textbf{99.843} & \cellcolor[rgb]{ .847,  .847,  .847}99.777 & \cellcolor[rgb]{ .847,  .847,  .847}99.469 & \cellcolor[rgb]{ .847,  .847,  .847}99.842 & \cellcolor[rgb]{ .847,  .847,  .847}99.834 \\
\hhline{|~|~|-|-|-|-|-|-|}
          & \%PF  & 3.3   & 3.8   & 4.4   & 1.2   & 4.2   & \textbf{4.9} \\
\hhline{--------}

\multirow{2}{*}{\emph{nrp-m4}} & \%Hyper  & \cellcolor[rgb]{ .847,  .847,  .847}99.840 & \cellcolor[rgb]{ .847,  .847,  .847}99.805 & \cellcolor[rgb]{ .847,  .847,  .847}\textbf{99.869} & \cellcolor[rgb]{ .847,  .847,  .847}99.571 & \cellcolor[rgb]{ .847,  .847,  .847}99.862 & \cellcolor[rgb]{ .847,  .847,  .847}99.825 \\
\hhline{|~|~|-|-|-|-|-|-|}
          & \%PF  & 4.1   & 3.4   & 5.1   & 1.7   & 4.8   & \textbf{5.3} \\
\hhline{--------}
%\bottomrule
\end{tabular}%
\caption{Results for anytime algorithms (excluding \emph{SPF}) within 60 seconds of time limit. Best percentages for every instance are marked in bold. \label{tab:anytime_60}}
\end{table*}%

As we can see in the results, every anytime algorithm can solve the \emph{nrp1} instance before the total time is reached. Moreover, every anytime method reaches more than the 99\% of the maximum hypervolume for all  NRP instances except for \emph{nrp2}, \emph{nrp4} and \emph{nrp5}, where the percentages are higher than 60\%, 67\% and 90\%, respectively. Nevertheless the results are quite similar, so we need to take three decimal numbers to show the differences between them. Notice that for \emph{nrp5}, algorithm \emph{AnyAugmecon(1,$\lambda$)} does not finish its execution because of an out of memory error, while \emph{AnyAugmecon(2,$\lambda$)} has the best results for that instance. In summary, it is clear that anytime algorithms behave as expected, because with a low number of the total solutions, but well-spread, we obtain a great percentage of the total hypervolume.

The best results are distributed among several algorithms, such as \emph{AnyAugmecon(1,$\lambda$)}, \emph{AnyAugmecon(2,$\lambda$)}, \emph{AnyHybrid}  and \emph{MixHT} for the best hypervolumes, and \emph{AnyAugmecon(2,$\lambda$)}, \emph{AnyHybrid} and \emph{MixSHT} for the best percentage of total solutions. 

We applied the Friedman test to the average hypervolume obtained by the methods to check if the differences are statistically significant. The result is a $p$-value of \numprint{3.187e-7}, that suggests a strong evidence that the performance of the algorithms is different. To find out the differences we do a post-hoc analysis using the Nemenyi multiple comparison test, available in an R-package called ~\emph{PMCMR}. The results are displayed in Table~\ref{tab:posthoc1}.

\begin{table}[!hb]
\scriptsize
\centering
\begin{tabular}{|c|c|c|c|c|c|}

%\hline 
\cline{2-6}    
\multicolumn{1}{c|} {} &
\begin{sideways} \emph{AnyAugmecon(2,$\lambda$) } \end{sideways}& 
\begin{sideways} \emph{AnyHybrid} \end{sideways}& 
\begin{sideways} \emph{AnyTchebycheff} \end{sideways}& 
\begin{sideways} \emph{MixHT} \end{sideways}& 
\begin{sideways} \emph{MixSHT} \end{sideways} \tabularnewline
\hline 
\emph{AnyAugmecon(1,$\lambda$) } & 0.15927 & 0.82732 & 1.2e-06 & 0.96213 & 0.00989 \tabularnewline
\hline 
\emph{AnyAugmecon(2,$\lambda$) } & - & 0.85074 & 0.03463 & 0.62413 & 0.92569 \tabularnewline
\hline
\emph{AnyHybrid} & - & - & 0.00048 & 0.99883 & 0.26315 \tabularnewline
\hline
\emph{AnyTchebycheff} & - & - & - & 8.3e-05 & 0.34184 \tabularnewline
\hline
\emph{MixHT} & - & - & - & - & 0.11337 \tabularnewline
\hline 
%\end{center}
\end{tabular}
\caption{Post-hoc analysis using Nemenyi multiple comparison test. \label{tab:posthoc1}}
\end{table}

As we can see, the only statistically significant differences  (at the 5\% significance level) are those of 
\emph{AnyTchebycheff} with the rest, excluding \emph{MixSHT}; and the pair \emph{AnyAugmecon(1,$\lambda$)} - \emph{MixHT}.
We use the Wilcoxon test to compare all the previous pairs and conclude that \emph{AnyAugmecon(1,$\lambda$)}, \emph{AnyAugmecon(2,$\lambda$)}, \emph{AnyHybrid} and \emph{MixHT} are better than \emph{AnyTchebycheff}, but there is no significant difference between \emph{AnyAugmecon(1,$\lambda$)} and \emph{MixSHT}. 

In conclusion, we can say that for these instances, \emph{AnyThebycheff} is worse than the others, but for the rest, there is no clear winner.

The main feature of the anytime methods is that they are able to increase the hypervolume very fast during the search process. To see this, we show in Figure~\ref{gnrp3} the curves for the anytime algorithms in instance \emph{nrp3} which provides a very good hypervolume for all of them after 10 seconds. However, there also exist differences at the beginning, being \emph{MixSHT} the best and \emph{AnyTchebycheff} the worst, in this case. In the supplementary material the reader can observe the progress in the hypervolume of all the anytime methods in all the NRP instances. 

\begin{figure*}[!ht]
\centering
\includegraphics [width=0.9\textwidth]{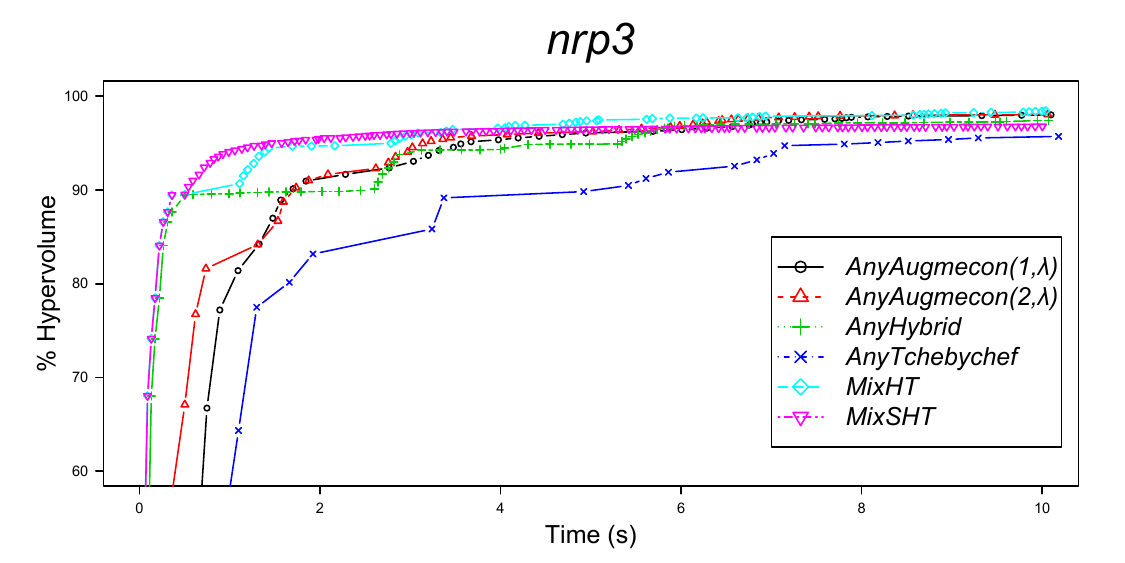}
\caption{Percentage of the total hypervolume for the six methods in the first 10 seconds for instance \emph{nrp3}.} \label{gnrp3}
\end{figure*}

\subsection{Answering RQ2: Traditional multi-objective against anytime algorithms}
\label{sec:computational_exact}

In this section we want to explore what is the real advantage of anytime methods compared to the classic multi-objective algorithms (described in Section~\ref{sec:exact_methods}). In order to do this, we run all the algorithms with a limited runtime: 60 seconds.
The results of the classic methods are displayed in Table~\ref{tab:exact_60}. Additional results of the classic methods can be found in the supplementary material. For each instance, we consider the total percentage of hypervolume and the total percentage of total solutions within 60 seconds. The $\varepsilon$-constraint method commented in Section \ref{sec:eps} had two variants, depending on which function we minimize. This is done including an input parameter \emph{obj} $\in \{1,2\}$. Considering the two approaches, we call \emph{Econst1} the one which uses one call to the solver at every iteration (with an ulterior filtering of weakly efficient solutions) and \emph{Econst2} the algorithm which solves two subproblems to obtain every non-dominated point. 
As we can see, every algorithm can solve the \emph{nrp1} instance before the total time is reached. On the other hand, some algorithms provide poor results in hypervolume or in the total number of solutions. As expected, algorithms \emph{Econst1} and \emph{Econst2} have the higher percentage of the total number of solutions found, but their hypervolume percentages are very poor because they are exploring the Pareto front in lexicographical order. They do not \emph{jump} in the objective space. Regarding the hypervolume, algorithm \emph{Tchebycheff} is clearly the best, maybe because of the structure of its level curves (see Section \ref{sec:exact_methods:tchebycheff}). We conclude from Table~\ref{tab:exact_60} that after 60 seconds the results for the hypervolume are around 70\% of the maximum hypervolume in the best cases, excluding \emph{nrp1}.

% Table generated by ExcelLaTeX from sheet 'tabla_exact_60'
\begin{table*}[!htb]
\small
%scriptsize
  \centering
  \begin{tabular}{|c|c|rrrrrrrr|}
\cline{3-10}
\multicolumn{1}{r}{} & 
\multicolumn{1}{l|}{ } & 
\multicolumn{1}{c|}{\begin{sideways}\textit{Econst1 (1)}\end{sideways}} & 
\multicolumn{1}{c|}{\begin{sideways}\textit{Econst1 (2)}\end{sideways}} & 
\multicolumn{1}{c|}{\begin{sideways}\textit{Econst2 (1)}\end{sideways}} & 
\multicolumn{1}{c|}{\begin{sideways}\textit{Econst2 (2)}\end{sideways}} & 
\multicolumn{1}{c|}{\begin{sideways}\textit{Augmecon (1,$\lambda$) } \end{sideways}} & 
\multicolumn{1}{c|}{\begin{sideways}\textit{Augmecon (2,$\lambda$)} \end{sideways}} & 
\multicolumn{1}{c|}{\begin{sideways}\textit{EHybrid}\end{sideways}} & 
\multicolumn{1}{c|}{\begin{sideways}\textit{Tchebycheff}\end{sideways}} \\
\hhline{----------}
    
\multirow{2}{*}{\emph{nrp1}} & \%Hyper & \cellcolor[rgb]{ .847,  .847,  .847}\textbf{100.0} & \cellcolor[rgb]{ .847,  .847,  .847}\textbf{100.0} & \cellcolor[rgb]{ .847,  .847,  .847}\textbf{100.0} & \cellcolor[rgb]{ .847,  .847,  .847}\textbf{100.0} & \cellcolor[rgb]{ .847,  .847,  .847}\textbf{100.0} & \cellcolor[rgb]{ .847,  .847,  .847}\textbf{100.0} & \cellcolor[rgb]{ .847,  .847,  .847}\textbf{100.0} & \cellcolor[rgb]{ .847,  .847,  .847}\textbf{100.0} \\
\hhline{|~|~|-|-|-|-|-|-|-|-|} 
          & \%PF  & \textbf{100.0} & \textbf{100.0} & \textbf{100.0} & \textbf{100.0} & \textbf{100.0} & \textbf{100.0} & \textbf{100.0} & \textbf{100.0} \\
\hhline{----------}

\multirow{2}{*}{\emph{nrp2}} & \%Hyper& \cellcolor[rgb]{ .847,  .847,  .847}27.0 & \cellcolor[rgb]{ .847,  .847,  .847}27.1 & \cellcolor[rgb]{ .847,  .847,  .847}25.1 & \cellcolor[rgb]{ .847,  .847,  .847}20.8 & \cellcolor[rgb]{ .847,  .847,  .847}23.2 & \cellcolor[rgb]{ .847,  .847,  .847}26.0 & \cellcolor[rgb]{ .847,  .847,  .847}44.0 & \cellcolor[rgb]{ .847,  .847,  .847}\textbf{67.2} \\
\hhline{|~|~|-|-|-|-|-|-|-|-|} 
          & \%PF  & 11.2  & \textbf{13.0} & 10.1  & 9.2   & 9.0   & 12.3  & 9.0   & 8.4 \\
\hhline{----------}

\multirow{2}{*}{\emph{nrp3}} & \%Hyper& \cellcolor[rgb]{ .847,  .847,  .847}31.1 & \cellcolor[rgb]{ .847,  .847,  .847}22.2 & \cellcolor[rgb]{ .847,  .847,  .847}26.9 & \cellcolor[rgb]{ .847,  .847,  .847}16.7 & \cellcolor[rgb]{ .847,  .847,  .847}27.6 & \cellcolor[rgb]{ .847,  .847,  .847}17.2 & \cellcolor[rgb]{ .847,  .847,  .847}\textbf{71.3} & \cellcolor[rgb]{ .847,  .847,  .847}68.7 \\
\hhline{|~|~|-|-|-|-|-|-|-|-|} 
          & \%PF  & \textbf{10.2} & 9.3   & 8.2   & 6.3   & 8.5   & 6.6   & 4.3   & 5.3 \\
\hhline{----------}

\multirow{2}{*}{\emph{nrp4}} & \%Hyper& \cellcolor[rgb]{ .847,  .847,  .847}12.4 & \cellcolor[rgb]{ .847,  .847,  .847}7.6 & \cellcolor[rgb]{ .847,  .847,  .847}10.2 & \cellcolor[rgb]{ .847,  .847,  .847}5.9 & \cellcolor[rgb]{ .847,  .847,  .847}12.9 & \cellcolor[rgb]{ .847,  .847,  .847}6.6 & \cellcolor[rgb]{ .847,  .847,  .847}64.8 & \cellcolor[rgb]{ .847,  .847,  .847}\textbf{67.7} \\
\hhline{|~|~|-|-|-|-|-|-|-|-|} 
          & \%PF  & 2.5   & 1.4   & 1.8   & 1.0   & \textbf{2.6} & 1.1   & 1.0   & 1.0 \\
\hhline{----------}

\multirow{2}{*}{\emph{nrp5}} & \%Hyper& \cellcolor[rgb]{ .847,  .847,  .847}60.9 & \cellcolor[rgb]{ .847,  .847,  .847}66.1 & \cellcolor[rgb]{ .847,  .847,  .847}71.0 & \cellcolor[rgb]{ .847,  .847,  .847}42.2 & \cellcolor[rgb]{ .847,  .847,  .847}* & \cellcolor[rgb]{ .847,  .847,  .847}54.6 & \cellcolor[rgb]{ .847,  .847,  .847}\textbf{74.6} & \cellcolor[rgb]{ .847,  .847,  .847}71.7 \\
\hhline{|~|~|-|-|-|-|-|-|-|-|} 
          & \%PF  & 22.4  & \textbf{49.7} & 30.5  & 29.5  & *     & 39.4  & 30.0  & 17.7 \\
\hhline{----------}

\multirow{2}{*}{\emph{nrp-e1}} & \%Hyper& \cellcolor[rgb]{ .847,  .847,  .847}24.0 & \cellcolor[rgb]{ .847,  .847,  .847}14.4 & \cellcolor[rgb]{ .847,  .847,  .847}18.9 & \cellcolor[rgb]{ .847,  .847,  .847}15.0 & \cellcolor[rgb]{ .847,  .847,  .847}19.8 & \cellcolor[rgb]{ .847,  .847,  .847}17.5 & \cellcolor[rgb]{ .847,  .847,  .847}70.9 & \cellcolor[rgb]{ .847,  .847,  .847}\textbf{71.8} \\
\hhline{|~|~|-|-|-|-|-|-|-|-|}
          & \%PF  & \textbf{7.7} & 3.4   & 5.5   & 3.6   & 5.9   & 4.4   & 2.2   & 2.9 \\
\hhline{----------}

\multirow{2}{*}{\emph{nrp-e2}} & \%Hyper& \cellcolor[rgb]{ .847,  .847,  .847}19.4 & \cellcolor[rgb]{ .847,  .847,  .847}11.6 & \cellcolor[rgb]{ .847,  .847,  .847}14.4 & \cellcolor[rgb]{ .847,  .847,  .847}12.5 & \cellcolor[rgb]{ .847,  .847,  .847}17.4 & \cellcolor[rgb]{ .847,  .847,  .847}14.7 & \cellcolor[rgb]{ .847,  .847,  .847}70.1 & \cellcolor[rgb]{ .847,  .847,  .847}\textbf{72.3} \\
\hhline{|~|~|-|-|-|-|-|-|-|-|}
          & \%PF  & \textbf{5.7} & 1.9   & 3.5   & 2.1   & 4.8   & 2.6   & 1.7   & 2.0 \\
\hhline{----------}

\multirow{2}{*}{\emph{nrp-e3}} & \%Hyper& \cellcolor[rgb]{ .847,  .847,  .847}35.6 & \cellcolor[rgb]{ .847,  .847,  .847}28.0 & \cellcolor[rgb]{ .847,  .847,  .847}29.3 & \cellcolor[rgb]{ .847,  .847,  .847}24.0 & \cellcolor[rgb]{ .847,  .847,  .847}26.8 & \cellcolor[rgb]{ .847,  .847,  .847}24.7 & \cellcolor[rgb]{ .847,  .847,  .847}70.7 & \cellcolor[rgb]{ .847,  .847,  .847}\textbf{72.1} \\
\hhline{|~|~|-|-|-|-|-|-|-|-|}    
        & \%PF  & \textbf{13.2} & 9.4   & 10.1  & 7.2   & 8.9   & 7.6   & 4.6   & 5.7 \\
\hhline{----------}

\multirow{2}{*}{\emph{nrp-e4}} & \%Hyper& \cellcolor[rgb]{ .847,  .847,  .847}34.6 & \cellcolor[rgb]{ .847,  .847,  .847}17.6 & \cellcolor[rgb]{ .847,  .847,  .847}25.6 & \cellcolor[rgb]{ .847,  .847,  .847}20.8 & \cellcolor[rgb]{ .847,  .847,  .847}27.6 & \cellcolor[rgb]{ .847,  .847,  .847}21.5 & \cellcolor[rgb]{ .847,  .847,  .847}70.5 & \cellcolor[rgb]{ .847,  .847,  .847}\textbf{71.8} \\
\hhline{|~|~|-|-|-|-|-|-|-|-|}
          & \%PF  & \textbf{13.5} & 4.4   & 8.5   & 5.6   & 9.5   & 5.9   & 3.7   & 4.8 \\
\hhline{----------}

\multirow{2}{*}{\emph{nrp-g1}} & \%Hyper& \cellcolor[rgb]{ .847,  .847,  .847}38.2 & \cellcolor[rgb]{ .847,  .847,  .847}44.7 & \cellcolor[rgb]{ .847,  .847,  .847}27.9 & \cellcolor[rgb]{ .847,  .847,  .847}36.6 & \cellcolor[rgb]{ .847,  .847,  .847}27.4 & \cellcolor[rgb]{ .847,  .847,  .847}38.1 & \cellcolor[rgb]{ .847,  .847,  .847}68.5 & \cellcolor[rgb]{ .847,  .847,  .847}\textbf{72.9} \\
\hhline{|~|~|-|-|-|-|-|-|-|-|}
          & \%PF  & \textbf{18.4} & 16.0  & 12.1  & 10.8  & 11.8  & 11.6  & 7.8   & 8.8 \\
\hhline{----------}

\multirow{2}{*}{\emph{nrp-g2}} & \%Hyper& \cellcolor[rgb]{ .847,  .847,  .847}42.8 & \cellcolor[rgb]{ .847,  .847,  .847}50.1 & \cellcolor[rgb]{ .847,  .847,  .847}29.7 & \cellcolor[rgb]{ .847,  .847,  .847}47.6 & \cellcolor[rgb]{ .847,  .847,  .847}32.9 & \cellcolor[rgb]{ .847,  .847,  .847}45.1 & \cellcolor[rgb]{ .847,  .847,  .847}69.7 & \cellcolor[rgb]{ .847,  .847,  .847}\textbf{73.9} \\
\hhline{|~|~|-|-|-|-|-|-|-|-|}
          & \%PF  & \textbf{25.6} & 13.7  & 15.8  & 12.5  & 18.3  & 11.5  & 8.1   & 8.9 \\
\hhline{----------}

\multirow{2}{*}{\emph{nrp-g3}} & \%Hyper& \cellcolor[rgb]{ .847,  .847,  .847}42.7 & \cellcolor[rgb]{ .847,  .847,  .847}48.4 & \cellcolor[rgb]{ .847,  .847,  .847}31.5 & \cellcolor[rgb]{ .847,  .847,  .847}40.5 & \cellcolor[rgb]{ .847,  .847,  .847}32.4 & \cellcolor[rgb]{ .847,  .847,  .847}41.6 & \cellcolor[rgb]{ .847,  .847,  .847}70.5 & \cellcolor[rgb]{ .847,  .847,  .847}\textbf{72.8} \\
\hhline{|~|~|-|-|-|-|-|-|-|-|}
          & \%PF  & \textbf{20.8} & 17.3  & 13.0  & 12.7  & 13.6  & 13.4  & 9.0   & 10.0 \\
\hhline{----------}

\multirow{2}{*}{\emph{nrp-g4}} & \%Hyper& \cellcolor[rgb]{ .847,  .847,  .847}54.2 & \cellcolor[rgb]{ .847,  .847,  .847}63.8 & \cellcolor[rgb]{ .847,  .847,  .847}37.8 & \cellcolor[rgb]{ .847,  .847,  .847}54.1 & \cellcolor[rgb]{ .847,  .847,  .847}42.6 & \cellcolor[rgb]{ .847,  .847,  .847}54.2 & \cellcolor[rgb]{ .847,  .847,  .847}70.5 & \cellcolor[rgb]{ .847,  .847,  .847}\textbf{73.4} \\
\hhline{|~|~|-|-|-|-|-|-|-|-|}
          & \%PF  & \textbf{32.5} & 26.4  & 18.5  & 18.6  & 22.1  & 18.7  & 12.9  & 14.4 \\
\hhline{----------}

\multirow{2}{*}{\emph{nrp-m1}} & \%Hyper& \cellcolor[rgb]{ .847,  .847,  .847}11.9 & \cellcolor[rgb]{ .847,  .847,  .847}9.2 & \cellcolor[rgb]{ .847,  .847,  .847}8.8 & \cellcolor[rgb]{ .847,  .847,  .847}8.9 & \cellcolor[rgb]{ .847,  .847,  .847}11.0 & \cellcolor[rgb]{ .847,  .847,  .847}12.3 & \cellcolor[rgb]{ .847,  .847,  .847}67.9 & \cellcolor[rgb]{ .847,  .847,  .847}\textbf{71.8} \\
\hhline{|~|~|-|-|-|-|-|-|-|-|}
          & \%PF  & \textbf{3.4} & 1.5   & 2.3   & 1.5   & 3.1   & 2.3   & 0.9   & 1.5 \\
\hhline{----------}
    
\multirow{2}{*}{\emph{nrp-m2}} & \%Hyper& \cellcolor[rgb]{ .847,  .847,  .847}12.0 & \cellcolor[rgb]{ .847,  .847,  .847}10.4 & \cellcolor[rgb]{ .847,  .847,  .847}8.4 & \cellcolor[rgb]{ .847,  .847,  .847}10.4 & \cellcolor[rgb]{ .847,  .847,  .847}10.8 & \cellcolor[rgb]{ .847,  .847,  .847}12.3 & \cellcolor[rgb]{ .847,  .847,  .847}67.2 & \cellcolor[rgb]{ .847,  .847,  .847}\textbf{72.1} \\
\hhline{|~|~|-|-|-|-|-|-|-|-|}
          & \%PF  & \textbf{3.7} & 1.3   & 2.3   & 1.3   & 3.2   & 1.8   & 0.8   & 1.2 \\
\hhline{----------}

\multirow{2}{*}{\emph{nrp-m3}} & \%Hyper& \cellcolor[rgb]{ .847,  .847,  .847}14.4 & \cellcolor[rgb]{ .847,  .847,  .847}10.7 & \cellcolor[rgb]{ .847,  .847,  .847}11.7 & \cellcolor[rgb]{ .847,  .847,  .847}9.4 & \cellcolor[rgb]{ .847,  .847,  .847}13.1 & \cellcolor[rgb]{ .847,  .847,  .847}12.1 & \cellcolor[rgb]{ .847,  .847,  .847}68.0 & \cellcolor[rgb]{ .847,  .847,  .847}\textbf{71.2} \\
\hhline{|~|~|-|-|-|-|-|-|-|-|}    
          & \%PF  & \textbf{4.0} & 2.9   & 3.0   & 2.4   & 3.5   & 3.5   & 1.4   & 2.1 \\
\hhline{----------}

\multirow{2}[2]{*}{\emph{nrp-m4}} & \%Hyper& \cellcolor[rgb]{ .847,  .847,  .847}16.4 & \cellcolor[rgb]{ .847,  .847,  .847}12.5 & \cellcolor[rgb]{ .847,  .847,  .847}11.2 & \cellcolor[rgb]{ .847,  .847,  .847}12.7 & \cellcolor[rgb]{ .847,  .847,  .847}13.9 & \cellcolor[rgb]{ .847,  .847,  .847}14.7 & \cellcolor[rgb]{ .847,  .847,  .847}68.0 & \cellcolor[rgb]{ .847,  .847,  .847}\textbf{71.7} \\
\hhline{|~|~|-|-|-|-|-|-|-|-|}
          & \%PF  & \textbf{5.5} & 2.3   & 3.4   & 2.4   & 4.4   & 3.1   & 1.5   & 1.9 \\
\hhline{----------}
%\bottomrule
\end{tabular}
\caption{Results for the classic multi-objective algorithms within 60 seconds of time limit. \label{tab:exact_60}}
\end{table*}

In order to answer RQ2, in Table~\ref{tab:compare_any_exact_60}, we compare the best classic algorithm against the worst anytime algorithm, in terms of hypervolume. We see that for all instances there are more than 22\% of improvement for anytime methods, except for \emph{nrp2} and \emph{nrp5}, where the best classic algorithm is better than the worst anytime, but not better than the best anytime. Interestingly, most of the best results for the classic exact algorithms are achieved with \emph{Tchebycheff} method, and most of the worst ones for anytime algorithms are with \emph{AnyTchebycheff} method. This fact shows that augmented Tchebycheff method works much better when it is used as anytime algorithm.
We see in Figures~\ref{nrp2c}, \ref{nrp4c}, \ref{nrp5c}, \ref{nrp-e4c} and \ref{nrp-m1c}, the progress in hypervolume of the best and worst classic exact and anytime algorithms for some of the NRP instances (in the supplementary material the progress of all the instances can be found).
 
As expected, all anytime algorithms do a much better job than classic algorithms to keep a well-spread set of solutions, as the progress in the hypervolume indicates.

% Table generated by Excel2LaTeX from sheet 'compare_exact_any_60'
\begin{table}[!ht]
  \scriptsize
  \centering
   \begin{tabular}{|c|cc|cc|c|}
\cline{2-6}
\multicolumn{1}{r|}{} & 
\multicolumn{1}{c|}{Best classic} & \%Hyper & 
\multicolumn{1}{c|}{Worst anytime} & \%Hyper & 
\% Differ. \\

\hline
    \emph{nrp2} & \textit{Tchebycheff} & 67.2  & \textit{AnyHybrid} & 60.2  & -7.0 \\
    \hline
    \emph{nrp3} & \textit{EHybrid} & 71.3  & \textit{AnyTchebycheff} & 99.4  & \textbf{28.1} \\
    \hline
    \emph{nrp4} & \textit{Tchebycheff} & 67.7  & \textit{MixSHT} & 90.4  & \textbf{22.7} \\
    \hline
    \emph{nrp5} & \textit{EHybrid} & 74.6  & \textit{AnyTchebycheff} & 67.8  & -6.8 \\
    \hline
    \emph{nrp-e1} & \textit{Tchebycheff} & 71.8  & \textit{AnyTchebycheff} & 99.7  & \textbf{27.9} \\
    \hline
    \emph{nrp-e2} & \textit{Tchebycheff} & 72.3  & \textit{AnyTchebycheff} & 99.6  & \textbf{27.3} \\
    \hline
    \emph{nrp-e3} & \textit{Tchebycheff} & 72.1  & \textit{AnyTchebycheff} & 99.8  & \textbf{27.7} \\
    \hline
    \emph{nrp-e4} & \textit{Tchebycheff} & 71.8  & \textit{AnyTchebycheff} & 99.8  & \textbf{28.0} \\
    \hline
    \emph{nrp-g1} & \textit{Tchebycheff} & 72.9  & \textit{AnyTchebycheff} & 99.9  & \textbf{27.0} \\
    \hline
    \emph{nrp-g2} & \textit{Tchebycheff} & 73.9  & \textit{AnyTchebycheff} & 99.9  & \textbf{26.0} \\
    \hline
    \emph{nrp-g3} & \textit{Tchebycheff} & 72.8  & \textit{AnyTchebycheff} & 99.9  & \textbf{27.1} \\
    \hline
    \emph{nrp-g4} & \textit{Tchebycheff} & 73.4  & \textit{AnyTchebycheff} & 99.9  & \textbf{26.5} \\
    \hline
    \emph{nrp-m1} & \textit{Tchebycheff} & 71.8  & \textit{AnyTchebycheff} & 99.4  & \textbf{27.6} \\
    \hline
    \emph{nrp-m2} & \textit{Tchebycheff} & 72.1  & \textit{AnyTchebycheff} & 99.4  & \textbf{27.3} \\
    \hline
    \emph{nrp-m3} & \textit{Tchebycheff} & 71.2  & \textit{AnyTchebycheff} & 99.5  & \textbf{28.3} \\
    \hline
    \emph{nrp-m4} & \textit{Tchebycheff} & 71.7  & \textit{AnyTchebycheff} & 99.6  & \textbf{27.9} \\
    \hline
    %\bottomrule
\end{tabular}
\caption{Comparing the approximated total percentage of hypervolume for the best classic algorithm versus the worst anytime algorithm. We omit instance \emph{nrp1} because its Pareto front is completely found by all methods. \label{tab:compare_any_exact_60} }
\end{table}%

\begin{figure*}[!ht]
\centering
\includegraphics [width=0.9\textwidth]{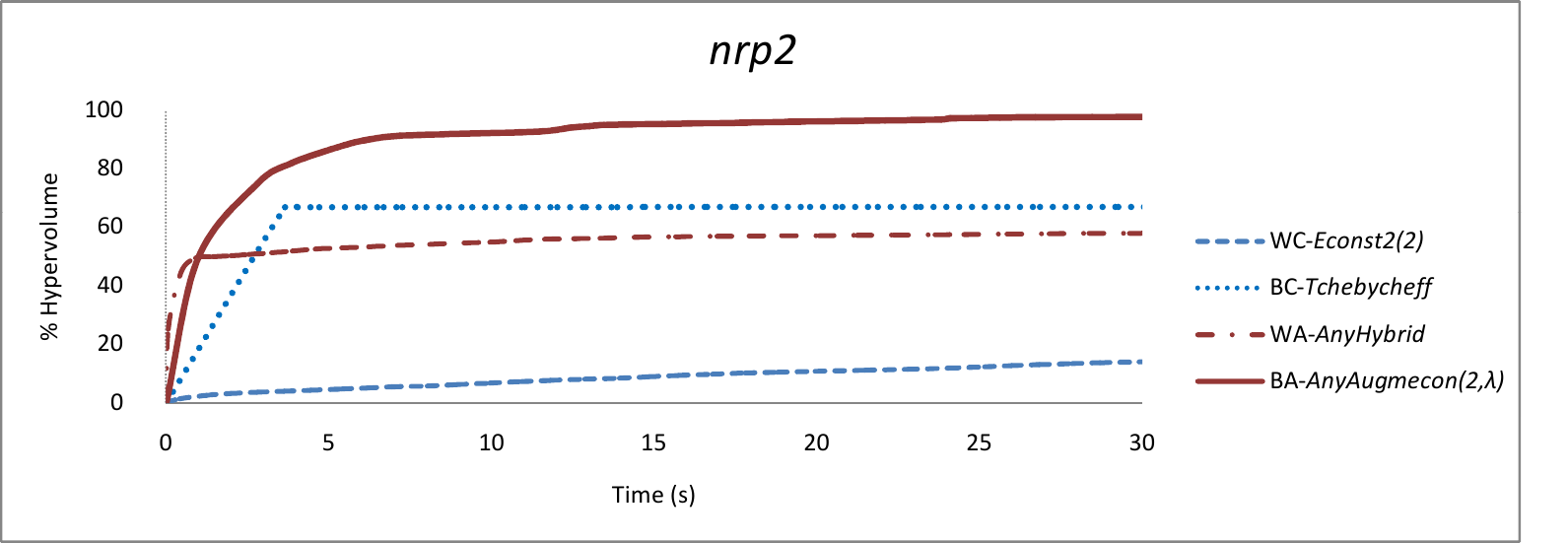}
\caption{Comparing worst classic exact (WC), best classic exact (BC), worst anytime (WA) and best anytime (BA) algorithms for instance  \emph{nrp2}.} \label{nrp2c}
\end{figure*}

\begin{figure*}[!ht]
\centering
\includegraphics [width=0.9\textwidth]{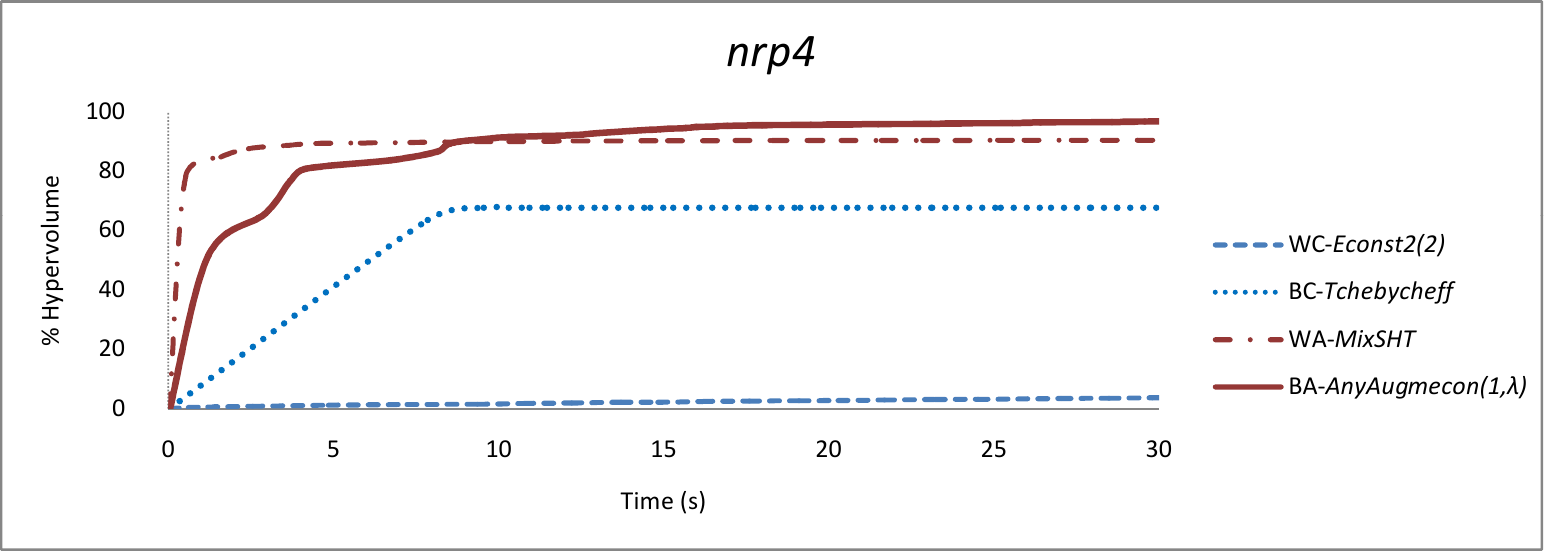}
\caption{Comparing worst classic exact (WC), best classic exact (BC), worst anytime (WA) and best anytime (BA) algorithms for instance  \emph{nrp4}.} \label{nrp4c}
\end{figure*}

\begin{figure*}[!ht]
\centering
\includegraphics [width=0.9\textwidth]{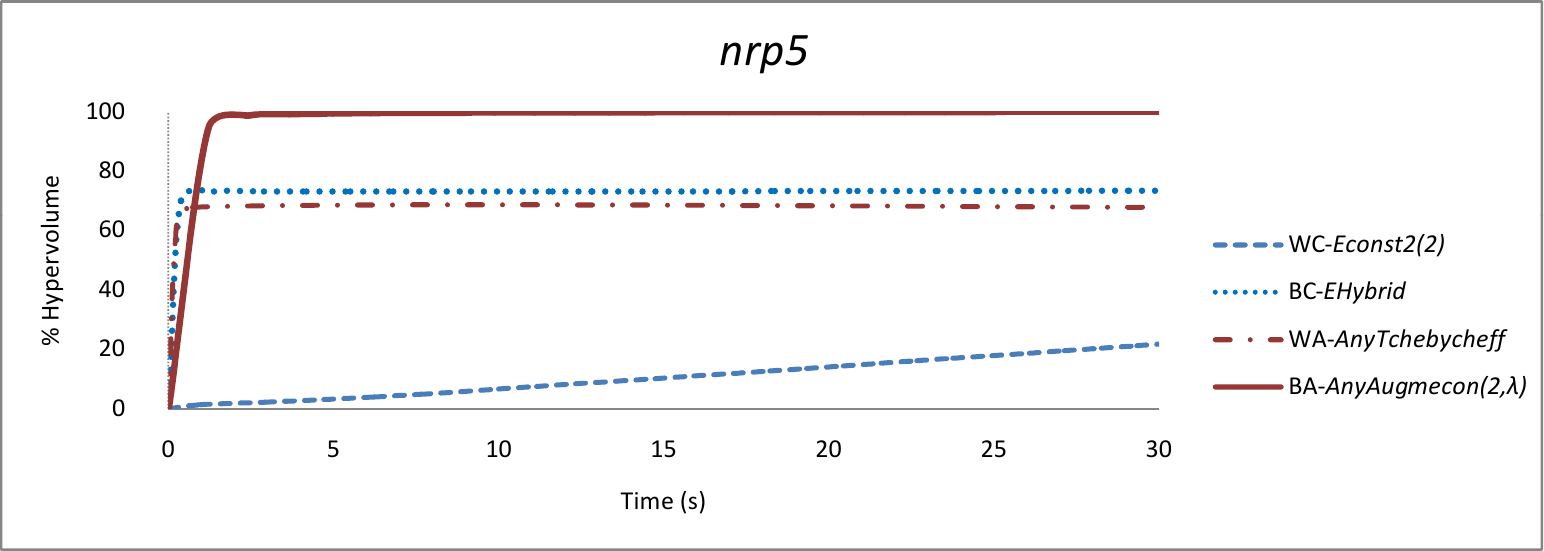}
\caption{Comparing worst classic (WC), best classic (BC), worst anytime (WA) and best anytime (BA) algorithms for instance  \emph{nrp5}.} \label{nrp5c}
\end{figure*}

\begin{figure*}[!ht]
\centering
\includegraphics [width=0.9\textwidth]{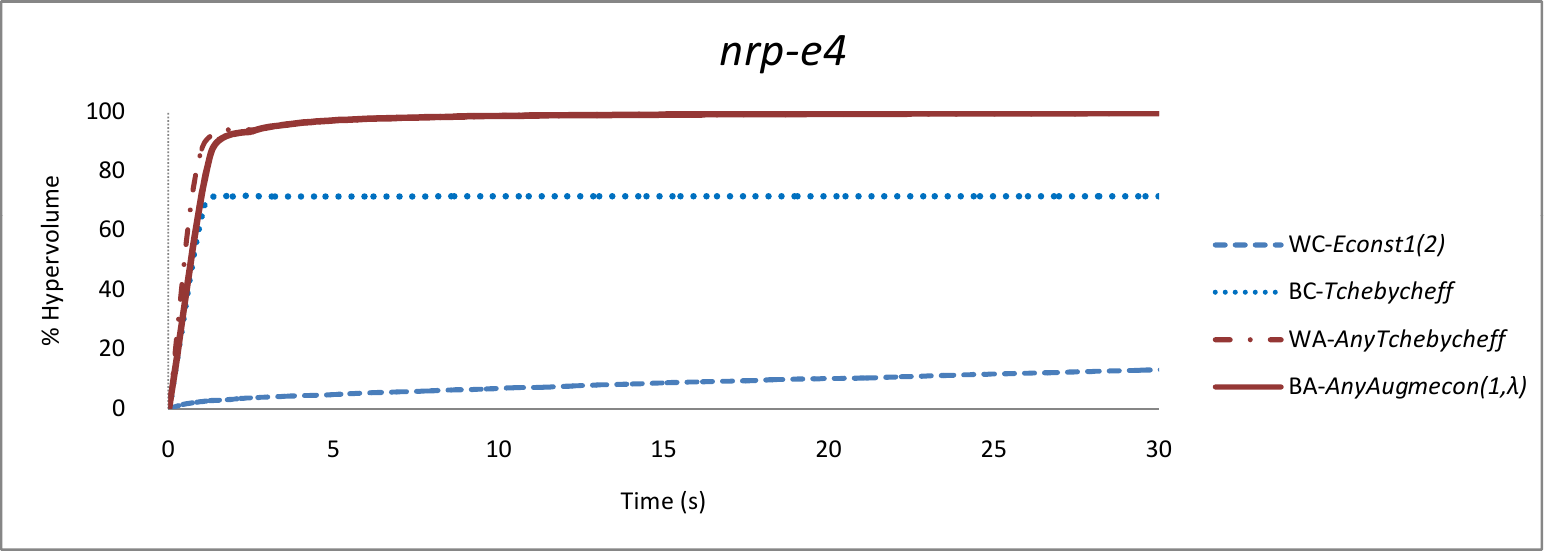}
\caption{Comparing worst classic exact (WC), best classic exact (BC), worst anytime (WA) and best anytime (BA) algorithms for instance  \emph{nrp-e4}.} \label{nrp-e4c}
\end{figure*}

\begin{figure*}[!ht]
\centering
\includegraphics [width=0.9\textwidth]{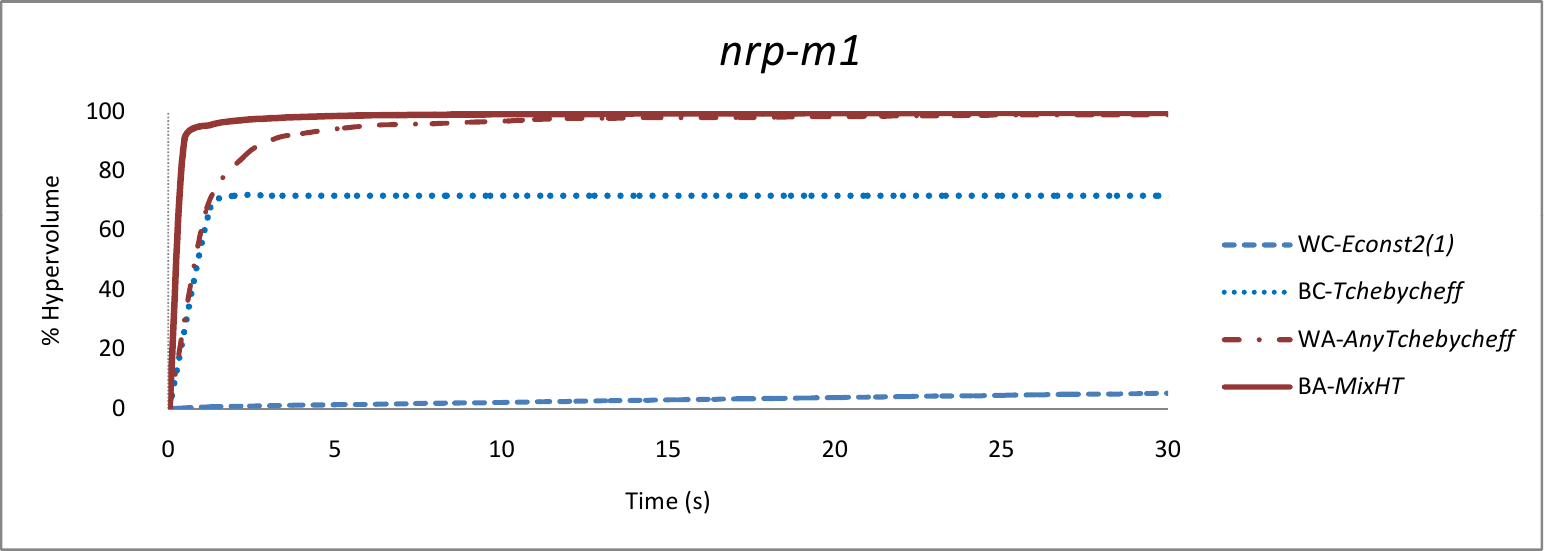}
\caption{Comparing worst classic exact (WC), best classic exact (BC), worst anytime (WA) and best anytime (BA) algorithms for instance  \emph{nrp-m1}.} \label{nrp-m1c}
\end{figure*}

\section{Discussion}
\label{sec:discussion}

In this section we discuss the connection between the results obtained in the previous section and the results in the literature, in particular, regarding the application of metaheuristic algorithms. We also analyze the utility of the proposed anytime methods for requirements engineering. 

\subsection{Results with metaheuristic algorithms}

The bi-objective Next Release Problem has been solved in the past using metaheuristic algorithms~\cite{aguila2016,Li2007,Li2010,Li2016,li2014robust,zhang2007multi}, the main reason being that the problem is NP-hard and exact methods require too much time. The work of Veerapen et al.~\cite{veerapen2015integer} is the most recent one claiming that an exact solution using ILP solvers is possible for this problem in a reasonable time, but they also use a Metaheuristic algorithm (NSGA-II) to compare the results with, showing that the metaheuristic algorithm is competitive with the Dichotomic search. In Section~\ref{sec:computational_anytime} we have shown that the anytime methods proposed here clearly beat the Dichotomic Search and are able to find the complete Pareto front if this is the desire of the user. Thus, we conclude that, for the sizes of the instances used in our experimental evaluation, anytime methods, proposed in this paper, should be clearly the preferred methods to find an appropriate (well-spread) set of efficient solutions for the bi-objective NRP. They have the advantage over the dichotomic search that all the non-dominated solutions are found (with enough time), not only the supported  solutions. They have also the advantage over the metaheuristic algorithms that all the solutions found are provably efficient (metaheuristics cannot guarantee that the solutions found are efficient) and they can be faster. In our previous report~\cite{chicano2016estrategias} it was clear that anytime methods outperform NSGA-II, GRASP and ACO, both in runtime and quality of solutions.

\subsection{Anytime methods in requirement engineering}

Regarding the use of anytime methods in Requirements Engineering, they are specially useful in the following scenarios:
\begin{itemize}
\item To check ``What if'' scenarios that allow the user to interactively try different values for the cost or value of the requirements in a short time. A slow method (like $\varepsilon$-constraint) is not appropriate for this purpose, since the user has to wait for the answer before checking a different scenario. Furthermore, the value and cost of the requirements are usually not precisely known, they are uncertain. Anytime algorithms can help to try different combinations of the requirements' parameters in a short time. This approach has been used in the past by Li et al.~\cite{li2017value}, and they conclude that the use of exact algorithms (like the proposed in this work) is important to avoid algorithmic uncertainty.
\item Sensitivity analysis and uncertainty, recently studies for the  problem by Li et al.~\cite{Li2016,li2014robust} require fast exact methods to find the solutions to the problem. Thanks to the use of anytime methods, this sensitivity analysis is possible in a short time (minutes) compared to the previous approaches that would require days of computation.
\item While the requirements selection is a problem to be solved every few months using a traditional waterfall methodology, in agile methodologies the sprints usually last for one or two weeks, and the selection of requirements (user stories) for a sprint is something done every one or two weeks. Thus, the time to solve the problem should be accordingly short compared to the duration of the sprint. A runtime of eight hours is too much time to make the selection. A few seconds or minutes, as the anytime methods require, is more appropriate.
\item When the number of requirements is in the order of tens or hundreds of thousands, finding the complete Pareto front is not viable in a reasonable time, but finding a set of a few well-spread solutions is possible using anytime algorithms.
\end{itemize}

When the selection of the requirements to implement in the next release does not need to be solved very often, non-anytime exact methods (like the ones proposed by Veerapen et al.~\cite{veerapen2015integer}) are also useful. They require a few hours to compute the Pareto front, but in these cases the algorithm to find the efficient solutions should be run every few months. Thus, the fact that the algorithms takes a few hours to compute the Pareto front is not a big issue for the software development team, and anytime methods have no clear advantages in these cases.

\section{Threats to validity}
\label{sec:threats}

%Las amenazas de validez de construcción se refieren a la relación entre teoría y observación. Se refiere al grado en que una prueba mide lo que dice que está midiendo. 
Construct validity concerns the relation between theory and observation. We use the hypervolume metric to assess the quality of the results. This quality is potentially subjective to the decision maker's opinion. Moreover, the hypervolume measures the convergence to the front and the spread of the solutions. Since in this paper the convergence of the solutions to the front is assured because the algorithms are exact, the hypervolume measures the spread of the solutions.

%La validez interna se refiere a las relaciones reales que se examinan
Internal validity is concerned with the causal relationships that are examined. We are working with exact algorithms, but their runtime is critical in our study and is subject to stochasticity due to the load of the  machines used for the experiments and the internal mechanisms of CPLEX. We used several runs for every NRP Instance and  statistical procedures to evaluate the results. The randomness of the process is mitigated with the high number of runs. On the other hand, we used the non-parametric Friedman test combined with a post-hoc analysis using the Nemenyi multiple comparison test to check the differences between the anytime algorithms. Thus, the conclusions are supported by statistical tests in order to mitigate the potential errors caused by stochasticity.

%La validez externa se refiere a la medida en que es posible generalizar los resultados
External validity concerns the possibility to generalize our results. The NRP instances used are varied in the number of variables and constraints, and also in the type of the constraints. We have used well-known benchmarks of instances. We were not able to compare with real-world instances, but we have a benchmark with realistic ones. The results in~\cite{veerapen2015integer} show that real-world instances are usually smaller than benchmark instances, so we think that our approach should be applicable also to real-world instances.

\section{Literature review}
\label{sec:review}

Many ranking, release planning and prioritization techniques have been defined, each one using a subset of the information collected for requirements~\cite{achimugu2014, Berander2005, Svahnberg2010,  thakurta2017}.
These methods may differ in the way  priorities are computed, in the scale of values
used to represent the resulting ordering, and in the accuracy of the results. They vary from those that do not use numerical data about requirement attributes, such as the MoSCoW method, the Top Ten ranking and the 100-point method, to more complex approaches that combine different steps, algorithms and software tools, such as 
 InSCo-Requisite \cite{aguila2016}, DRank \cite{Shao2017} and DMGame \cite{Kifetew2017}. InSCo-Requisite    
defines a process flow across three stages: gathering candidate requirements, finding solutions to the problem using metaheuristic algorithms, and analyzing the solutions found. DRank makes use of machine learning techniques to guide the user preferences elicitation in the prioritization process and extracts requirements dependencies  from  requirements models. DMGame exploits  game  elements: AHP (analytic hierarchy process) and genetic algorithms in an iterative prioritization process. 

%%%%% RElated works based on SBSE approach

The Next Release Problem has as goal to meet the customer’s needs, minimizing development effort and maximizing customers satisfaction. It was originally proposed by Bagnall et al.~\cite{bagnall2001next} at customer level and by van den Akker et al.~\cite{Akker2005} at requirements level.
The first approach did not give any value property to each requirement, it is estimated according to the weight or the client importance. The goal of the problem is to select the subset of the requirements to be satisfied that maximize the satisfaction of the involved clients without overcoming the budget taking as reference the estimated efforts. In the second problem, a value is assigned to individual requirements to model their importance. % model the importance or value is assigned to individual requirements. 
This way, the individual profit for each requirement can be estimated. The goal of the problem is to select the subset of the requirements that maximize their values without exceeding the cost bound. This formulation is more accurate to current software engineering development approaches where selection has to be done at feature level. The bi-objective NRP was formulated by Zhang et al.~\cite{zhang2007multi} naming it Multi-Objective Next Release Problem (MONRP). In this case, the upper bound of the cost is lifted and that constraint is transformed into a second objective. Then, the decision-maker is presented with a set of solutions which are all efficient in the Pareto sense.

%%%%% Requirements interactions
Another point to be considered in the problem definition is requirements interaction~\cite{Karlsson1997RE}, that is, constraints among the requirements that must be considered. Some works prioritize the interactions to the value-cost criterion for requirement triage \cite{Davis2003, Saliu2007}, that is, interactions represent a stronger constraint than the resources. 
These interactions were unified and classified as strong and weak (functional, value based) \cite{Carlshamre2001}, but until 2002 they were not totally formalized \cite{Carlshamre2002}. The complete list of interactions, including exclusion and time-value dependencies, appeared later \cite{Li2007,Li2010,Akker2008}. Interactions  are  constraints that should be represented in the problem formulation. Precedence relations are first represented by a graph \cite{bagnall2001next}, combination relations and function interactions also are represented as a graph in \cite{Ngo-The2004} and \cite{Sagrado2011}, respectively.

%%%% SBSE techniques single objective
In the literature, several search techniques showed promising results when only one objective is managed. Some examples are  hill climbing \cite{bagnall2001next}, simulated annealing \cite{bagnall2001next, Sagrado2010}, integer linear programming \cite{bagnall2001next, Akker2005}  genetic algorithms \cite{Greer2004,Ruhe2003}, ant colony optimization \cite{de2011ant, Sagrado2010} and approximate backbone based multilevel algorithm \cite{xuan2012solving}.

%%%% NSGDP NEW PARAGRAHPH
The work recently published by Li et al.~\cite{li2017value} deserves a special mention. They developed a decision support framework for the Next Release Problem to manage algorithmic and requirement uncertainty. Using a conflict graph to model the mutual exclusion between requirements, and considering the possibility of partial satisfaction for the stakeholders, they applied the Nemhauser-Ullmann algorithm, which is a dynamic programming method, to solve the NRP. Their algorithm, called NSGDP, cannot be compared to our algorithms for two reasons. The first one is that the Nemhauser-Ullmann algorithm can not deal with constraints in the requirements, and we have prerequisites (a kind of requirement) in all our classic instances. The second one is that the formulation of the Next Release Problem we use does not consider partial satisfaction of the stakeholders.

%%%% SBSE techniques multiobjective
The multi-objective approach finds a set of non-dominated solutions. Most of the works that manage MONRP apply the cost-value approach (minimal cost and maximal client satisfaction) in several ways: as an interplay between requirements and implementation constraints \cite{Saliu2007}, considering two objectives (cost and value) \cite{zhang2007multi}, using different measures of fairness \cite{Finkelstein2009}, applying several algorithms based on genetic inspiration (such as NSGA-II, MOCell and PAES) \cite{durillo2011study,Durillo2009}, applying multiobjective ant-colony algorithms \cite{Sagrado2015}, using differential evolution (a kind of evolutionary algorithm) \cite{Chaves-Gonzalez2015}, and using grey wolf optimization algorithm and clustering approach \cite{Masadeh2018}. However, other objectives have also been considered, such as
client dissatisfaction, risk or urgency, \cite{Ngo2008,Pitangueira2017}. 

There are some works that propose combining search techniques with human preferences \cite{Babar2015,aguila2016,Ferreira2017}, learning algorithms \cite{Araujo2017}, statistical methods to deal with uncertainty \cite{Li2016,li2014robust} and AHP \cite{Tonella2013}.

%%%%%%%%%%%%%%%%%%%% Exact techniques  ILP

Integer linear programming (ILP) had been applied from the very beginning to NRP even before the name NRP was coined. Jung \cite{Jung1998} used this method to reduce the complexity of AHP to large instances of the problem.
Bagnall et al.~\cite{bagnall2001next}, who named the problem, had used exact techniques to solve a linear programming relaxation of the problem, in addition to greedy and hill climbing algorithms. They concluded that, despite the results, there was scope for further development on both heuristic and exact techniques, as has been demonstrated along the more than fifteen last years. 

As Bagnall et al.~\cite{bagnall2001next} said, linear programming solutions proved to be sufficient on small problem instances but required a long time for larger problems.
ILP was also used in a release planning tool that managed requirements interactions  \cite{Carlshamre2002} and stakeholder’s opinions for release planning \cite{Ruhe2005}.  

An extended  ILP technique that manages the list of requirements,  requirements' interactions, requirements' projected revenue, and requirements' resource claim per development team was proposed later to support software vendors in determining the next release \cite{Akker2005,Akker2008}. 
Two integer ILP models that integrate requirement selection into software release planning have been successfully used to minimize project duration in the first model and to maximize revenues and calculate an on-time-delivery project schedule
\cite{Li2007,Li2010}. A reconsideration of
ILP for the single-objective formulation of the problem and its integration within the $\epsilon$-constraint method has also been used to address the MONRP \cite{veerapen2015integer}. 
Exact approaches are unappealing when the number of requirements or interactions grows up because of large run times. Iterated applications of ILP (solving  a  series  of  single  objective  subproblems) are used to generate the exact Pareto front obtaining very fast results on smaller instances of the problem but can take several
hours for larger, more complex instances 
\cite{veerapen2015integer}.

None of the previous work using exact techniques focused on anytime methods. This paper makes a contribution to the line of research using exact ILP-based methods to solve the bi-objective formulation of the Next Release Problem. We propose five anytime methods that improve the state-of-the-art in the problem by finding a well-spread set of solutions in a few seconds for instances with up to several thousands of requirements.

\section{Conclusions and future work}
\label{sec:conclusions}

Many optimization problems in Software Engineering can be modeled as multi-objective optimization problems. This is the case of the bi-objective Next Release Problem used here. Finding the whole Pareto front for these problems is time consuming and unnecessary in most of the cases, since the decision maker just needs a few solution well-spread in the objective space to take the decision. We propose here some exact algorithms to find a well-spread set of solutions at anytime from the beginning of the search. We have seen that, in practice, for the Next Release Problem, they obtain a set of well-spread solutions in the objective front within a few seconds, while the complete front require several hours of computation for the instances used. We claim that this kind of algorithm (anytime) should be the preferred ones by the decision makers in Software Engineering, since they allow them to play with different parameters and have exact answers in seconds. In the literature of the Next Release Problem, however, most of the works use metaheuristic algorithms, that cannot guarantee that efficient solutions are found.

We have worked here with the bi-objective Next Release Problem, but the same idea can be applied to other Software Engineering Problems as future work. The main key ingredient for a successful application of anytime algorithms is an efficient exact method to find the efficient solutions. Regarding the Next Release Problem, there are other variants  where the value or satisfaction are not certain or depend on the presence/absence of other requirements. These variants would require a different, more complex, formulation to be solved with our anytime algorithms that can be addressed in future work. Other lines of future work include solving largest instances, probably combining exact methods and heuristics, improving the anytime algorithms, and extending them to more than two objectives.

\section*{Acknowledgements}

This research has been partially funded by the Spanish Ministry of Economy and Competitiveness (MINECO) and the European Regional Development Fund (FEDER), under contracts TIN2014-57341-R (moveOn project), TIN2015-71841-REDT (SEBASENet Excellence Network), TIN2016-77902-C3-3-P (PGM-SDA II project) and TIN2017-88213-R (6city project). The authors also acknowledge the funds of the University of Málaga for the EXHAURO Project (PPIT.UMA.B1.2017/07).

%\bibliographystyle{amsplain}
%\bibliography{bibliografia_NRP}

\providecommand{\bysame}{\leavevmode\hbox to3em{\hrulefill}\thinspace}
\providecommand{\MR}{\relax\ifhmode\unskip\space\fi MR }
\providecommand{\MRhref}[2]{%
  \href{http://www.ams.org/mathscinet-getitem?mr=#1}{#2}
}
\providecommand{\href}[2]{#2}

\end{document}